\documentclass[onecolumn]{pasj01_arxiv}
\usepackage{comment}
\usepackage{url}

\usepackage{lineno}
\begin{document} 

\Received{2021/02/24}
\Accepted{2021/10/17}
\Published{}

\title{Nobeyama 45 m Local Spur CO survey. I. Giant molecular filaments and cluster formation in the Vulpecula OB association}

\author{Mikito \textsc{Kohno}\altaffilmark{1}$^{*}$}%

\author{Atsushi \textsc{Nishimura}\altaffilmark{2,{5, 9}}}
\email{kohno@nagoya-p.jp}
\email{mikito@a.phys.nagoya-u.ac.jp}

\author{Shinji \textsc{Fujita}\altaffilmark{2}}
\author{Kengo \textsc{Tachihara}\altaffilmark{3}}
\author{Toshikazu \textsc{Onishi}\altaffilmark{2}}
\author{Kazuki \textsc{Tokuda}\altaffilmark{2,4}}
\author{Yasuo \textsc{Fukui}\altaffilmark{3}}
\author{Yusuke \textsc{Miyamoto}\altaffilmark{4}}

\author{Shota \textsc{Ueda}\altaffilmark{2}}
\author{{Ryosuke \textsc{Kiridoshi}\altaffilmark{2} }}
\author{Daichi \textsc{Tsutsumi}\altaffilmark{3}}
\author{Kazufumi \textsc{Torii}\altaffilmark{5}}
\author{Tetsuhiro \textsc{Minamidani}\altaffilmark{4,6}}
\author{Kazuya \textsc{Saigo}\altaffilmark{4}}
\author{Toshihiro \textsc{Handa}\altaffilmark{7,8}}
\author{Hidetoshi \textsc{Sano}\altaffilmark{4} }


\altaffiltext{1}{Astronomy Section, Nagoya City Science Museum, 2-17-1 Sakae, Naka-ku, Nagoya, Aichi 460-0008, Japan}
\altaffiltext{2}{Department of Physical Science, Graduate School of Science, Osaka Prefecture University, 1-1 Gakuen-cho, Naka-ku, Sakai, Osaka 599-8531, Japan}
\altaffiltext{3}{Department of Physics, Graduate School of Science, Nagoya University, Furo-cho, Chikusa-ku, Nagoya, Aichi 464-8602, Japan}
\altaffiltext{4}{National Astronomical Observatory of Japan (NAOJ), National Institutes of Natural Sciences (NINS), 2-21-1 Osawa, Mitaka, Tokyo 181-8588, Japan}
\altaffiltext{5}{Nobeyama Radio Observatory, National Astronomical Observatory of Japan (NAOJ), National Institutes of Natural Sciences (NINS), 462-2, Nobeyama, Minamimaki, Minamisaku, Nagano 384-1305, Japan}
\altaffiltext{6}{Department of Astronomical Science, School of Physical Science, SOKENDAI (The Graduate University for Advanced Studies), 2-21-1, Osawa, Mitaka, Tokyo 181-8588, Japan}
\altaffiltext{7}{Graduate School of Science and Engineering, Kagoshima University, 1-21-35 Korimoto, Kagoshima, Kagoshima 890-0065, Japan}
\altaffiltext{8}{Amanogawa Galaxy Astronomy Research Center (AGARC), Kagoshima University, 1-21-35 Korimoto, Kagoshima, Kagoshima 890-0065, Japan}
\altaffiltext{9}{{Institute of Astronomy, The University of Tokyo, 2-21-1, Osawa, Mitaka, Tokyo 181-0015, Japan}}


\maketitle

\begin{abstract}
We have {performed} new large-scale $^{12}$CO, $^{13}$CO, {and} C$^{18}$O $J=$1--0 observations toward the Vulpecula OB association ($l \sim \timeform{60D}$) as part of the Nobeyama 45 m Local Spur CO survey project.
Molecular clouds {are distributed} over {$\sim 100$} pc, {with} local peaks {at the}  Sh 2-86, Sh 2-87, and Sh 2-88 high-mass star-forming regions in the Vulpecula complex. 
The molecular gas is associated with the Local {Spur}, which corresponds to {the nearest inter-arm {region} located} between the Local {Arm} and {the} Sagittarius {Arm}.
{We discovered new giant molecular filaments (GMFs) in Sh 2-86{,}  with a {length} of $\sim 30$ pc, {width} of $\sim 5$ pc, and molecular {mass} of $\sim 4\times 10^4\ M_{\odot}$}.
We also found that Sh 2-86 {contains} the three velocity components at 22, 27, and 33 km s$^{-1}$. These clouds {and GMFs} are likely to be physically associated with Sh 2-86 because they have high $^{12}$CO $J =$ 2--1 to $J =$ 1--0 intensity ratios and {coincide} with the infrared dust emission. The open cluster NGC 6823 exists at the {common} intersection of these clouds. 
{We argue that the multiple cloud interaction scenario, including GMFs, can explain cluster formation in the Vulpecula OB association}. 

%
\end{abstract}

\section{Introduction}
Giant molecular clouds (GMCs) are the {sites} of high-mass star and cluster formation (e.g., \cite{2003ARA&A..41...57L,2007ARA&A..45..565M}).
 Their formation and evolution processes have been studied in the Milky Way and Local Group Galaxies (e.g., \cite{2007prpl.conf...81B,2010ARA&A..48..547F, 2014prpl.conf....3D}).
{Galactic-plane} CO surveys by single-dish radio telescopes {have} revealed {the} large-scale {distribution} of GMCs in the Milky Way  (e.g., \cite{1991ARA&A..29..195C, 2015ARA&A..53..583H}). 
{The} {GMCs} exist not only in spiral arms but also {in} inter-arm regions called {'spurs', 'branches' or 'bridges' } (e.g., \cite{1980ApJ...239L..53C,1986ApJ...305..892D}).
{In particular, \citet{2014A&A...568A..73R} reported that filamentary structures{, which they} named giant molecular filaments (GMFs) exist {in} the inter-arm regions of the Milky Way. 
Long filamentary clouds like GMFs often include {the} dense molecular gas {that forms} massive stars and clusters (e.g., \cite{2010ApJ...719L.185J, 2020A&A...641A..53W}). 
Thus, their formation and evolution have been investigated by numerical simulations (e.g., \cite{2017MNRAS.470.4261D}) and observations (e.g., \cite{2013A&A...559A..34L,2014ApJ...797...53G,2015ApJ...815...23Z,2015MNRAS.450.4043W,2016A&A...590A.131A, 2019A&A...622A..52Z}). }
Recently, {a} new spur {located} between the Local {Arm} and Sagittarius {Arm,} has been revealed by parallax measurements of maser spots {from the very-long-baseline interferometry} (VLBI: \cite{2016SciA....2E0878X}) and {of OB-type stars from} Gaia data \citep{2018A&A...616L..15X,2021arXiv210100158X}. 
{In this paper, we refer to this spur structure as a "Local Spur" (see Figure 1 in \cite{2016ApJ...823...77R}).}

{GMCs are essential targets {for studying} star formation in galaxies, {but} it is not yet clear {what processes are responsible for the formation of massive stars and clusters in inter-arm regions}. The Local Spur is the best site {for investigating these processes,} because {it is the inter-arm region nearest to} the solar system. 
In this paper, we present high-resolution CO observations toward the Vulpecula OB association{, with the goal of investigating the star-formation processes} in a GMC {in this} inter-arm {region} in the solar neighborhood.
This paper is constructed as follows: {Section 2} introduces the Vulpecula OB association{,} {Section} 3 presents the datasets{, and} {Section} 4 gives the results{, including the} physical parameters of {the} molecular {clouds. In Section 5,} we discuss the NGC 6823 {cluster-formation} scenario, and in Section 6{,} we summarize this paper.}

\section{{The Vulpecula} OB association {in the Local Spur}}
\begin{figure*}[h]
\begin{center} 
 \includegraphics[width=16cm]{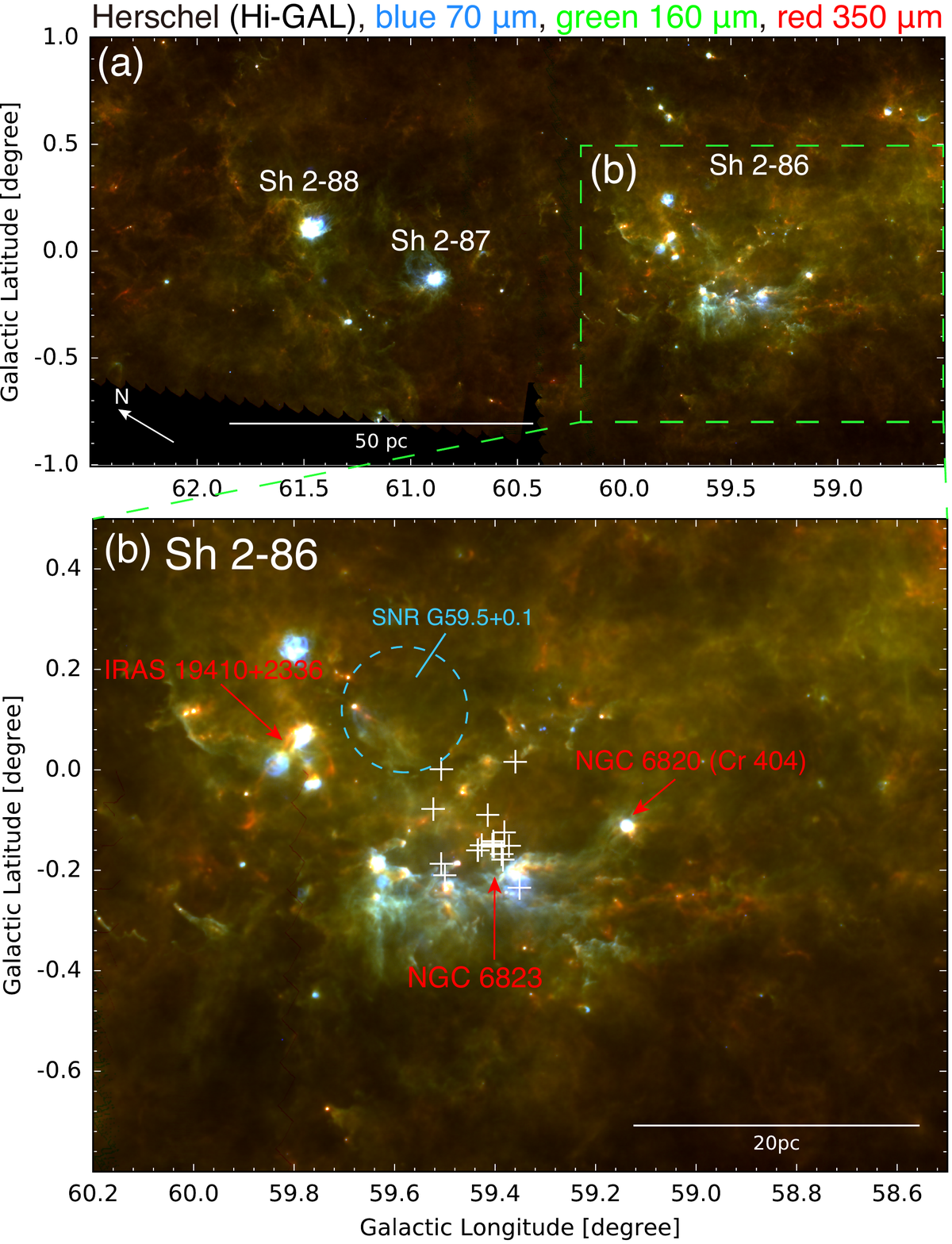}
\end{center}
\caption{(a) {{Herschel}} three-color composite image of the Vulpecula OB association. Blue, green, and red {represent} the {Herschel}/PACS 70 $\>\mu$m, {Herschel}/PACS 160 $\>\mu$m, and {Herschel}/SPIRE 350 $\>\mu$m {distributions, respectively} (\cite{2016A&A...591A.149M}). (b) Close up image of Sh 2-86 {(see also Supplemental Figure 5 in \cite{2018ARA&A..56...41M})}. The white crosses indicate the OB-type stars in NGC 6823 identified by \citet{1995ApJ...454..151M}. The blue dotted circle shows the position of SNR G59.5+0.1 \citep{2019JApA...40...36G}.}
\label{Vul_HiGAL}
\end{figure*}

\begin{table*}[h]
\tbl{Basic parameters of each H\,\emissiontype{II} region in Vulpecula}{
\begin{tabular}{cccccccccc}
\hline
\multicolumn{1}{c}{Name} & $l$  & $b$& Luminosity &  Earliest & Age & References\\
 & [$^\circ$] &[$^\circ$]& [$L_{\odot}$]   &  Spectral Type & &\\
\hline
Sh 2-86 & 59.36  & {$-0.18$} &$ \sim 10^4${$^*$} &O7V {$^\ddagger$} &$3 \pm 1$ Myr{$^\ddagger$} &[1,2,3]\\
Sh 2-87 &   60.88  & {$-0.13$} & $5.0 \times 10^4${$^*$}   &B0 &3-4 Myr & [4,5, 6]\\
Sh 2-88 &  61.47  & {0.10} & $1.3 \times 10^5${$^\dag$}  &O8.5V-O9.5V &$\sim 2$ Myr &[4, 6, 7, 8, 9]\\
\hline
\end{tabular}}
\label{Ostar_Vul}
\begin{tabnote}
[1] \citet{2000AcA....50..113P}, [2] \citet{1995ApJ...454..151M}, [3] \citet{2008ApJ...681..428C}, [4] \citet{2003ARA&A..41...57L}, [5] \citet{2003A&A...401..185C}, [6] \citet{2007ApJ...659..459S}, [7] \citet{2000A&A...360.1107D},[8] \citet{2009A&A...502..559M}, [9] \citet{2002A&A...395..955C} \\
{The assumed distances are 2.3 kpc$^*$  and 2.0 kpc$^\dag$. $\ddagger$ The values are {taken} from the cluster in NGC 6823.}
\end{tabnote}
\end{table*}

{The} Vulpecula OB association (hereafter Vul OB1) is {a} high-mass star-forming region, which was first cataloged by \citet{1953ApJ...118..318M}.
It {is located} {at} $l \sim \timeform{60D}$ in the Galactic plane, and {it} is {a} {nearby ($<$3 kpc)} massive molecular cloud complex in the Milky Way ({see Table 3 in \cite{2018ARA&A..56...41M}}).
The Vul OB1 GMC {contains} the three H\,\emissiontype{II} regions: Sh 2-86, Sh 2-87, and Sh 2-88 \citep{1959ApJS....4..257S}.
{Figures \ref{Vul_HiGAL}(a) and \ref{Vul_HiGAL}(b)} show  three-color composite {images} obtained {with} the {{Herschel}} space telescope \citep{2010A&A...518L...1P} as part of the {{{Herschel}} infrared Galactic Plane Survey} (Hi-GAL) project \citep{2010PASP..122..314M}.
{The 70 $\mu$m (blue) and 160 $\mu$m (green) emissions trace the warm dust, whereas the 350 $\mu$m (red) emission traces the cold dust (e.g., \cite{2013A&A...554A..42R})}.
Sh 2-86 has diffuse cold dust emission, {whereas} Sh 2-87 and Sh 2-88 {display compact} bright emission {regions}. 
The basic parameters of {these} three H\,\emissiontype{II} regions are summarized in Table \ref{Ostar_Vul}. 

\subsection{Sh 2-86}
{Figure \ref{Vul_HiGAL} (b) shows a close-up image of Sh 2-86 obtained by {Herschel} (see also Figure 2 in \cite{2010A&A...518L.100M} and {Supplemental Figure 5 in \cite{2018ARA&A..56...41M}}).}
Sh 2-86 is an H\,\emissiontype{II} region that {includes the} open cluster NGC 6823 (=Cr 405: \cite{1931AnLun...2....1C}), which has been studied by optical observations for more than 60 years (e.g., \cite{1957SvA.....1..822B,1971A&A....10..270E,1979JRASC..73...74T, 1981Ap&SS..75..465S, 1988AJ.....96.1389S, 1992AJ....103..197G, 1999A&AS..136..313S,2010ApJ...715.1132O,2012MNRAS.420...10M}). 
\citet{1995ApJ...454..151M} identified 17 OB-type stars as members of this open cluster (Table \ref{OBstar_NGC6823}). 
Previous studies have estimated the ages of cluster members to lie within the range 1-5 Myr \citep{2012MNRAS.419.1887R, 2000AcA....50..113P,2004MNRAS.353..991K}.
{The embedded cluster NGC 6820 ($=$Cr 404, IRAS 19403+2258: \cite{2008A&A...489.1129B}), which corresponds to a compact infrared peak in the {{Herschel}} image, {is located} on the western side of Sh 2-86 at $(l, b) \sim (\timeform{59.14D}, \timeform{-0.11D})$}.

\begin{table*}[h]
\tbl{Lists of OB-type stars in NGC 6823}{
\begin{tabular}{cccccccccc}
\hline
\multicolumn{1}{c}{Name} & Galactic Longitude  & Galactic Latitude & Spectral Type  \\
& [degree]  & [degree]  & \\
\hline
  &59.50&	$-0.21$&	B1 V  \\
	&59.51&$-0.19$&B1.5 III  	\\
	&59.35&$-0.24$&B1 V      	\\
Erick 93	&59.43& $-0.16$&	B1 V      	\\
Hoag 8	&59.38&$-0.18$&	B1.5 V    	\\
Hoag 9	&59.43&$-0.15$&	B0.5 V    	\\
Hoag 6	&59.39&$-0.17$&	B0.5 III  	\\
Hoag 2	&59.40&$-0.15$&	O7 V((f)) 	\\
Sharp d	&59.41&$-0.15$&	B0.5 V    	\\
Sharp e	&59.40&	$-0.15$&	B0.5 V    	\\
HD 344775&59.52&	$-0.078$	&	B1 III    	\\
Sharp n  	&59.40&	$-0.14$&	O9.5Ia    	\\
HD 344783&59.37&	$-0.15$&	O9.5Ia    	\\
Hoag 10	&59.38&	$-0.12$&	B2 V      	\\
Hoag 7	&59.41&	$-0.090$	&	B1.5 V    	\\
HD 344776  &59.51&$0.0015$	&	B0.5 Ia   	\\
	&59.36&$0.016$&	B1.5 V    	\\
\hline
\end{tabular}}
\label{OBstar_NGC6823}
\begin{tabnote}
References: \citet{1995ApJ...454..151M}\\
\end{tabnote}
\end{table*}

{The hyper-compact H\,\emissiontype{II} region IRAS 19410+2336 (GAL 059.7+00.1) {is found} at $(l, b) \sim (\timeform{59.78D}, {\timeform{+0.06D}})$ {about {15 pc} from the cluster NGC 6823} (e.g., \cite{2008ApJ...681..428C}). 
{Energetic outflows and maser sources have been} reported in this source (e.g., \cite{2000A&AS..143..269S, 2004ApJ...615..832B,2012A&A...545A..51R,2008A&A...489..229M}).}
 \citet{1992AJ....103..931T} discovered the supernova remnant (SNR) G59.5+0.1, {with a} radius {of \timeform{15'}} and {an} age {of} $10^3$ - $10^4$ yr \citep{2005ChJAA...5..165X}.
\citet{2012A&A...543A..24X} argued that the G59.5+0.1 progenitor {may have} induced star formation around {the} SNR.

\subsection{{Sh 2-87 and Sh 2-88}}
Sh 2-87 is a massive star-forming region located {at} $(l, b) \sim (\timeform{60.88D}, {\timeform{-0.13D}})$, and {it is} composed of compact H\,\emissiontype{II} {regions}.
 \citet{1989ApJ...345..268B} reported {a} molecular outflow from the central embedded source. 
 The  Lyman continuum flux derived from {observations with the Very Large Array} is $1.9 \times 10^{47}$ s$^{-1}$, which corresponds to a B0-type exciting star \citep{1989ApJ...345..268B}.

 Sh 2-88 is a compact H\,\emissiontype{II} region located {at} $(l, b) \sim (\timeform{61.47D}, {\timeform{0.10D}})$, and {it is} excited by an O8.5V-O9.5V star \citep{2000A&A...360.1107D, 2002A&A...395..955C}.
The compact H\,\emissiontype{II} {regions} Sh 2-88A and Sh 2-88B {have} been studied by optical, infrared, and radio observations {since} the 1970s {as places where compact H\,\emissiontype{II} regions {interact} with molecular clouds} (e.g., \cite{1974A&A....30...67L, 1977A&A....59..215P, 1978A&A....70...19D,1981ApJ...250..200E}).

\clearpage
\subsection{{Star formation scenario and distance {to} the Vulpecula complex}}
{A propagating} star formation scenario {has been} discussed {for} Vul OB1 \citep{1986A&A...167..157T, 2001A&A...374..682E} based on  \citet{1977ApJ...214..725E}.
On the other hand, \citet{2010ApJ...712..797B} excluded the sequential-star-formation scenario because {evolutionary} stage of the YSO population is {homogeneously distributed} in the Vul OB1 GMC.
{In addition, it} has been suggested that the clusters in Sh 2-87 and  Sh 2-88 may have originated in star formation {triggered} by clump-clump collisions \citep{2008ApJ...680..446X,2009ApJ...705..468H, 2010ApJ...719.1813H}. 

{These} previous studies suggest that the H\,\emissiontype{II} regions and clusters in {the} GMC have {a} {common} distance of 2.0-2.3 kpc (e.g., \cite{2010ApJ...712..797B}).
In this paper, we {adopt} 2.0 kpc, which is the mean {of the parallactic distances} measured by VLBI {to the} masers in G059.47-00.18 (1.87 kpc: \cite{2016SciA....2E0878X}) and G59.7+0.1 (2.16 kpc:\cite{2009ApJ...693..413X}).
Figure \ref{survey_area}(b) shows the position of Vul OB1 in the Milky Way based on trigonometric parallax measurements \citep{2019ApJ...885..131R,2020arXiv200203089V}.
 
{The molecular} clouds in Vul OB1 were discovered in the 1980s {using a} 1.2 m radio telescope, but it {their} relationship {to} star-formation activity {was not clear} because of the low {\timeform{8'}} angular resolution {of this {study}} (e.g., \cite{1985ApJ...297..751D}). Studies of star formation were carried out only in {the individual} H\,\emissiontype{II} {regions} and open clusters, {whereas the entire} GMC in Vul OB1 {has not yet been studied}.
Therefore, we performed large-scale high-resolution {($\sim$ \timeform{20"})} CO observations to investigate the relationship of {the} molecular clouds {to} star formation in the Vul OB1 GMC.

\section{{Data sets}}
\begin{table*}[h]
\tbl{Observational properties of data sets.}{
\begin{tabular}{cccccccccc}
\hline
\multicolumn{1}{c}{Telescope/Survey} & Line & Receiver & HPBW &Effective   &  Velocity & RMS noise$^{\dag}$ & References \\
& && &Resolution &  Resolution & level &\\
\hline
Nobeyama 45-m/ &$^{12}$CO $J=$ 1--0 \footnotemark[]  & FOREST & \timeform{14"} & $\sim$ \timeform{20"}  & 0.5 $\>$km s$^{-1}$ & $\sim 0.70$ K  & This work\\
Local Spur &$^{13}$CO $J=$ 1--0\footnotemark[] & FOREST & \timeform{15"} &$\sim$  \timeform{21"}  &  0.5 $\>$km s$^{-1}$& $\sim 0.30$ K  & This work \\
&C$^{18}$O $J=$ 1--0\footnotemark[] &FOREST & \timeform{15"} &$\sim$ \timeform{21"}  &  0.5 $\>$km s$^{-1}$& $\sim 0.30$ K & This work \\
\hline
Osaka Pref. 1.85-m/Galactic Plane &$^{12}$CO $J=$ 2--1 \footnotemark[]  & & \timeform{2.7'} &{$\sim$ \timeform{3.4'}}  & 0.08 $\>$km s$^{-1}$ & $\sim 0.50$ K  & [1,2, 3]\\
\hline
\hline
Telescope/Survey & Band  & Detector &  & Resolution & References & & \\
\hline
{Herschel}/Hi-GAL   & 70 $\>\mu$m & PACS & & $\sim$\timeform{6"} & [4,5] & \\
{Herschel}/Hi-GAL   & 160 $\>\mu$m & PACS &  & $\sim$\timeform{12"} & [4,5]  &\\
{Herschel}/Hi-GAL   & 350 $\>\mu$m  & SPIRE & & $\sim$\timeform{24"} &  [4,6]  &\\
\hline
\end{tabular}}\label{obs_param}
\begin{tabnote}
\footnotemark[$\dag$] The value of rms noise levels are {for the smoothed (in space and velocity)} data sets. \\
References [1] \citet{2013PASJ...65...78O},[2] \citet{2015ApJS..216...18N}, [3] {\citet{2020arXiv201200906N}} [4] \citet{2016A&A...591A.149M}, [5] \citet{2010A&A...518L...2P}, [6] \citet{2010A&A...518L...3G}
\end{tabnote}
\end{table*}

\subsection{The Nobeyama 45 m Local Spur CO survey project: $^{12}$CO, $^{13}$CO, and C$^{18}$O $J=$1--0 observations}
\begin{figure*}[h]
\begin{center} 
 \includegraphics[width=17cm]{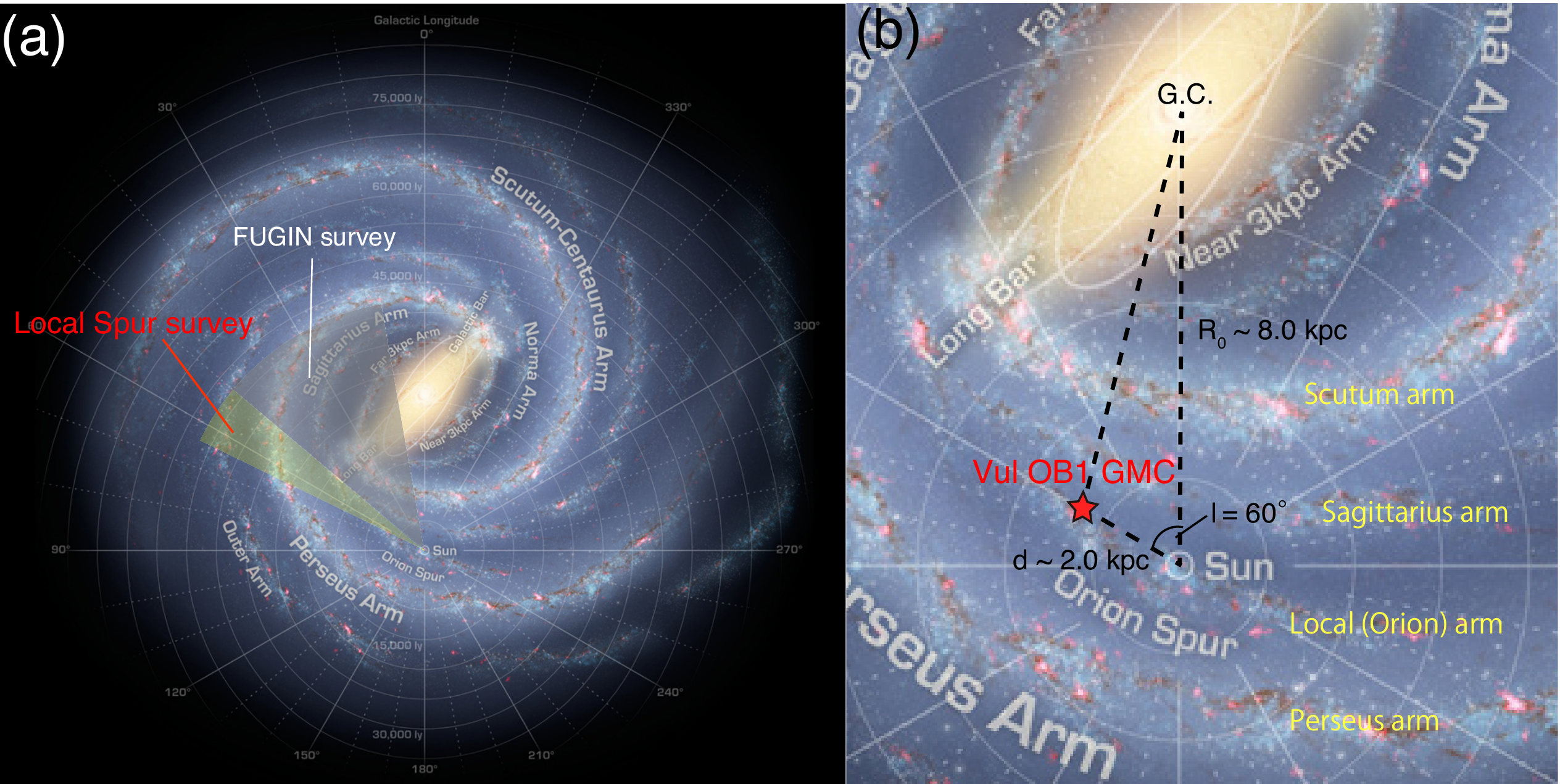}
\end{center}
\caption{Top-view of the Milky Way (NASA/JPL-Caltech/ESO/R. Hurt). (a) The gray and yellow shadows show the survey areas of FUGIN and the Local Spur CO survey project, respectively. (b)  The star symbol indicates the position of Vul OB1. The distances to the Galactic Center ($R_0$) and to Vul OB1 ($d$) are adopted from the VLBI astrometry {results: $R_0$ ($\sim 8.0$ kpc) was obtained by the averaged value of  \citet{2019ApJ...885..131R} and \citet{2020arXiv200203089V}, and $d$ ($\sim 2.0$ kpc) was obtained by the averaged value of Xu et al. (2009, 2016).}}
\label{survey_area}
\end{figure*}

We carried out CO $J=$ 1--0 observations toward the Vul OB1 GMC from December 2018 to March 2019 using the Nobeyama 45 m telescope. 
Our data {were} obtained as part of the Local Spur CO survey (CG181017: PI. A. Nishimura).
{Figures} \ref{survey_area} (a) and \ref{survey_area} (b) show the survey area and {a} close-up image of Vul OB1 GMC, superposed on {a} face-on view of the Milky Way. 
{The surveyed area is} $l=\timeform{50D}$-$\timeform{64D}$,  $b=\timeform{-1D}$-$\timeform{+1D}$ {, and we used} the on-the-fly (OTF){ scan-mapping} mode \citep{2008PASJ...60..445S}. 
We observed the Vul OB1 GMC {simultaneously} in $^{12}$CO (115.271 GHz), $^{13}$CO (110.201 GHz), and C$^{18}$O $J=$ 1--0 (109.782 GHz).
This CO survey is {an extension} of the FUGIN CO Galactic plane survey (\cite{2017PASJ...69...78U, 2019PASJ...71S...2T, 2015EAS....75..193M, 2018PASJ...70S..50K,2018PASJ...70S..51T,2018PASJ...70S..42N, 2018PASJ..tmp..121T,2019ApJ...872...49F};{\cite{2019PASJ...71S...1S}},\cite{2019PASJ...71S...2T, 2020arXiv200110693K, 2020arXiv200805939N};{\cite{2020PASJ...72...43N}}).
More detailed information will be {presented} in the project overview paper {(Nishimura et al. in preparation)}.
The {half-power beam width (HPBW)} of the 45 m telescope is \timeform{14"} in $^{12}$CO and \timeform{15"} in $^{13}$CO and {C$^{18}$O}. 
The effective beam size convolved {with a} the Bessel-Gaussian function is \timeform{20"} in $^{12}$CO and \timeform{21"} in $^{13}$CO and {C$^{18}$O}.
The front end was {the} four-beam, dual-polarization, sideband-separating (2SB) FOur-beam REceiver System on the 45 m Telescope (FOREST: \cite{2016SPIE.9914E..1ZM, 2019PASJ...71S..17N}).
The back-end system was a Spectral Analysis Machine for the 45 m telescope (SAM 45: \cite{Proc..2011}){,} with 4096 channels, which is the same as the ALMA ACA correlator \citep{2012PASJ...64...29K}.
The frequency bandwidth and resolution were 31.25 MHz and 15.26 kHz in each window, respectively, using the {spectral-window} mode\footnote{\url{https://www.nro.nao.ac.jp/~nro45mrt/html/prop/status/Status_latest.html#FOREST_SpW}}.
We utilized {a} chopper wheel to convert the raw data to antenna-temperature {units} ($T_a^*$) \citep{1976ApJS...30..247U, 1981ApJ...250..341K}.
We derived the scaling factor {by} converting $T_a^*$ to the main beam temperature ($T_{\rm mb}$) {and comparing it} with the FUGIN data {for} the standard source W51A \citep{2019PASJ..tmp...46F}.
The pointing accuracy was checked {to be} within \timeform{2"}-\timeform{3"} by observing the 43 GHz SiO maser source IRAS 19252+2201 $(l , b)= ({\timeform{56.61D}, \timeform{2.47D}}$) every 90 min using the H40 high electron mobility transistor (HEMT) receiver. 
The data {were} smoothed {using a two-dimensional} Gaussian {kernel} function with {the full width at half maximum of \timeform{36"}} to achieve the final resolution of \timeform{40"}. 
In this paper, we analyzed the final cube with the grid size of $(l, b, v)=(\timeform{10"}, \timeform{10"}, 1.0\ {\rm km\ s}^{-1})$.
The {root-mean-square} (r.m.s) noise levels are $\sim 0.70$ K,  $\sim 0.30$ K, and $\sim 0.30$ K in $^{12}$CO,  $^{13}$CO, and C$^{18}$O{, respectively}.

\subsection{{Osaka Prefecture University 1.85-m radio telescope: CO $J=$2--1 Galactic plane survey}}
We utilized $^{12}$CO $J=$2--1 data obtained {with} the 1.85-m radio telescope installed {at the} Nobeyama Radio Observatory and operated by Osaka Prefecture University  \citep{2013PASJ...65...78O,2015ApJS..216...18N,2020arXiv201200906N}.
The HPBW is \timeform{2.7'}, and the front-end is a 2SB SIS mixer receiver \citep{2017PASJ...69...91H}.
The back-end was {a} digital spectrometer having 16384 channels. 
The bandwidth and frequency resolutions were 1 GHz and 61 kHz, which correspond to 250 km s$^{-1}$ and 0.08 km s$^{-1}$ {in} velocity space{, respectively}.
The calibration was performed by observing {Orion KL as a} standard source \citep{2015ApJS..216...18N}\footnote{\url{http://www.astro.s.osakafu-u.ac.jp/~nishimura/Orion/}}. The uncertainty {in the} data is $\sim 10\%$.
More detailed information about the 1.85-m radio telescope is {provided by} \citet{2013PASJ...65...78O} and \citet{2020arXiv201200906N}. 
The effective resolution of the cube data is $\sim \timeform{3.4'}$.
We analyzed the final 3D cube having a voxel {resolution} of $(l, b, v) = (\timeform{60"},\timeform{60"}, 0.079 $ km s$^{-1}$).

\subsection{The Hi-GAL project: {Herschel} far-infrared archival data {for} the Galactic plane}
We also utilized the dust-continuum data {obtained} from the {{Herschel}} Space Observatory \citep{2010A&A...518L...1P}. 
The public DR1 fits data {were} taken from the VIALACTEA web page\footnote{http://vialactea.iaps.inaf.it/vialactea/eng/index.php}{, which is} part of Hi-GAL {survey} project \citep{2010PASP..122..314M, 2010A&A...518L.100M, 2016A&A...591A.149M}.
The far-infrared 70 and 160 $\>\mu$m images were obtained {with} the Photodetector Array Camera and Spectrometer (PACS).
 The 350 $\>\mu$m data were obtained {with} the Spectral and Photometric Imaging REceiver (SPIRE: \cite{2010A&A...518L...3G}). 
We summarize the basic parameters of {the datasets} in Table \ref{obs_param}.

\section{Results}
\subsection{Large-scale CO velocity distributions in the Local {Spur}}

\begin{figure*}[h]
\begin{center} 
 \includegraphics[width=16cm]{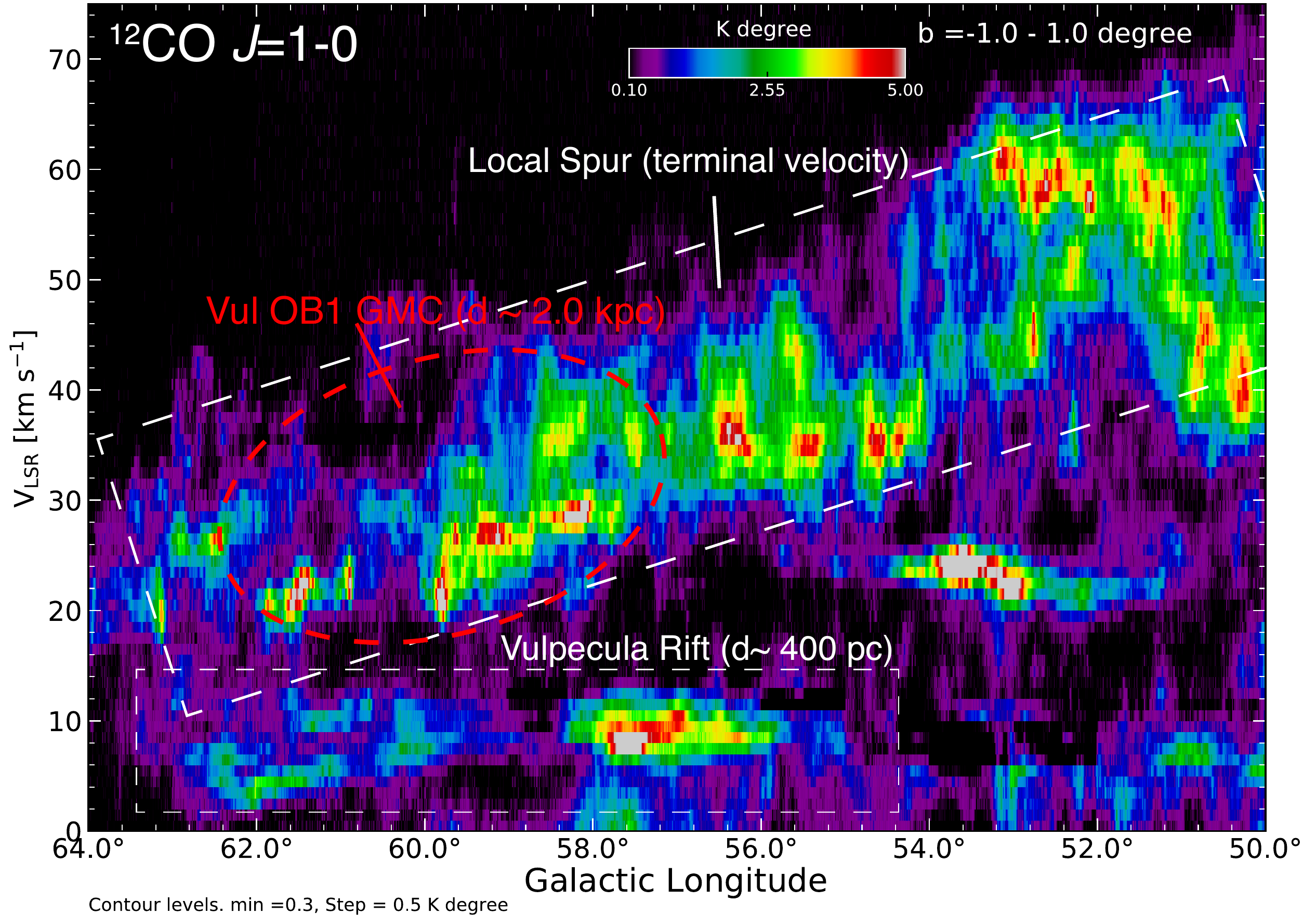}
\end{center}
\caption{ The large-scale $^{12}$CO $J=$ 1--0 longitude-velocity diagram obtained with the Nobeyama 45 m telescope. White dotted rectangles in the velocity ranges $0$- $+15$ km s$^{-1}$ and $15$-$70$ km s$^{-1}$ show the Vulpecula Rift and Local Spur, respectively. The red dotted circle indicates the Vul OB1 GMC.}
\label{Spur_largelv}
\end{figure*}
Figure \ref{Spur_largelv} shows the $^{12}$CO longitude-velocity diagram {for the entire} survey area.
We find mainly two components {in the velocity range} $0$-- $+15$ km s$^{-1}$ and 15--70 km s$^{-1}$ in this longitude range.
The former is the Vulpecula Rift, which {contains} local clouds {in} the solar neighborhood with {distances} of  $\sim 400$ pc (e.g., \cite{1985ApJ...297..751D, 1987ApJ...322..706D}). 
The latter{, which is distributed} diagonally {in} the longitude-velocity diagram, is the Local Spur{, which corresponds} to the inter-arm region between the Local {Arm} and Sagittarius {Arm} \citep{2016ApJ...823...77R, 2016SciA....2E0878X}.
This velocity range also corresponds to the terminal velocity in the first quadrant {of the Galaxy} (e.g., \cite{1978A&A....63....7B,2016ApJ...831..124M}).
The GMC associated with Vul OB1 is part of the {Local Spur in} the velocity range 15-40 km s$^{-1}$. We also point out another velocity cloud around $(l, v) \sim (\timeform{53D}, 23\ {\rm km\ s}^{-1})$. This cloud corresponds to GMF 54.0-52.0 \citep{2014A&A...568A..73R}. 
{GMF 54.0-52.0 includes dense gas \citep{2020A&A...641A..53W} and shows signatures of massive star formation in the N115 infrared bubble reported by previous studies \citep{2014A&A...569A..36X,2016A&A...595A..49Z}.}
A detailed analysis of this cloud will be {presented} in a separate paper.

\clearpage
\subsection{CO spatial distributions in Vul OB1}
\begin{figure*}[h]
\begin{center} 
 \includegraphics[width=15cm]{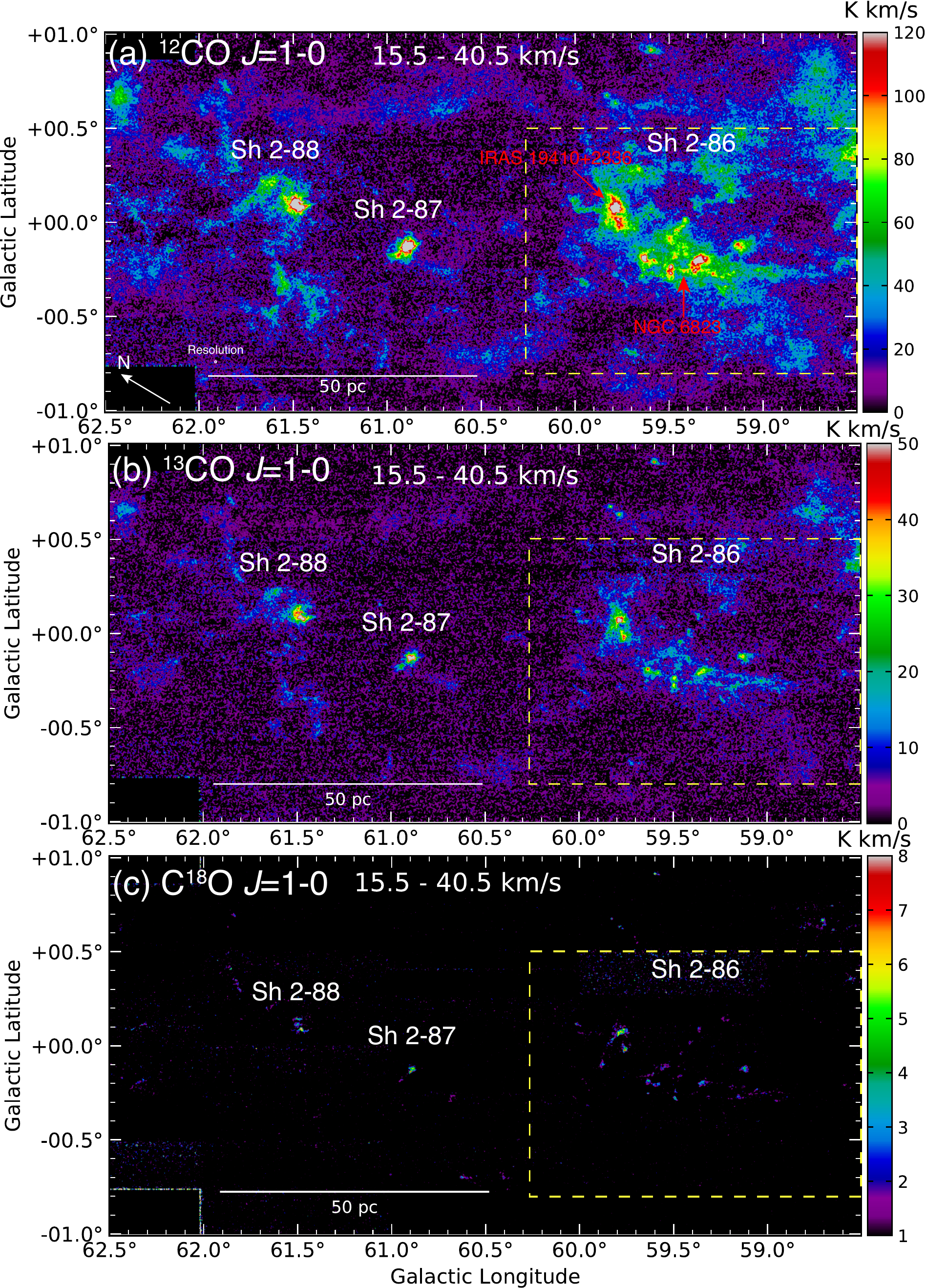}
\end{center}
\caption{Integrated intensity maps of (a) $^{12}$CO,  (b) $^{13}$CO, and (c) C$^{18}$O $J=$1--0 for Vul OB1. {The yellow dotted boxes show the area presented in the velocity-channel maps in {Figure \ref{12COch}}. {The map resolution} after convolution is \timeform{40"}. {Pixels of the C$^{18}$O map are blanked if the line intensity is less than 1.0 K.} }}
\label{integ}
\end{figure*}
Figure \ref{integ} presents the integrated intensity maps of (a) $^{12}$CO, (b) $^{13}$CO, and (c) C$^{18}$O. 
The molecular clouds have peaks in the three H\,\emissiontype{II} regions.
The intensity peaks in Sh 2-86 correspond to IRAS 19410+2336 and NGC 6823.
Sh 2-87 and Sh 2-88 have peaks in the {centers} of the H\,\emissiontype{II} regions.
The $^{13}$CO molecular gas {is distributed within} dense regions of $^{12}$CO.
The C$^{18}$O emission is hardly detected in {the} Vul OB1 GMC, and {it is} localized to {a small area within $\sim 1$ pc from} Sh 2-86, Sh 2-87,  and Sh 2-88.
 In this paper, we {have} carried out more detailed analyses toward Sh 2-86, where {the} molecular gas {is distributed} over $\sim 30$ pc.

{Figure \ref{12COch} shows} $^{12}$CO velocity channel maps {of} Sh 2-86. 
We find {a} filamentary cloud (Filament A) {that is} elongated from east to west {in the velocity range} 21.5-25.5 km s$^{-1}$, and {the} CO peaks correspond to the infrared source IRAS 19410+2336.
In {the velocity range} 23.5-26.5 km s$^{-1}$, we discovered {a} filamentary cloud (Filament B) {that extends over} {$\sim 40$ pc} from north to south.
We also find {a} filamentary cloud (Filament C) {in the velocity range} 26.5-29.5 km s$^{-1}$. 
Filament C has {a} length of {$\sim 30 $ pc}, and {it is} extended {in} the direction of Galactic longitude.
The CO peaks correspond to NGC 6823 and {to} the embedded cluster NGC 6820. 
{We point out that Filament B and C are likely to be comparable to {the} GMFs in the Milky Way reported by \citet{2014A&A...568A..73R}.}

On the other hand, {there is also a} round-shaped cloud  in the velocity range 31.5-34.5 km s$^{-1}$ {that is distributed} over $\sim 10$ pc around the {cluster NGC 6823}.
We also present $^{13}$CO and C$^{18}$O velocity channel maps of Sh 2-86 in {the} Appendix (see {Figures \ref{13COch1} and \ref{C18Och1}}).

\begin{figure*}[h]
\begin{center} 
 \includegraphics[width=14cm]{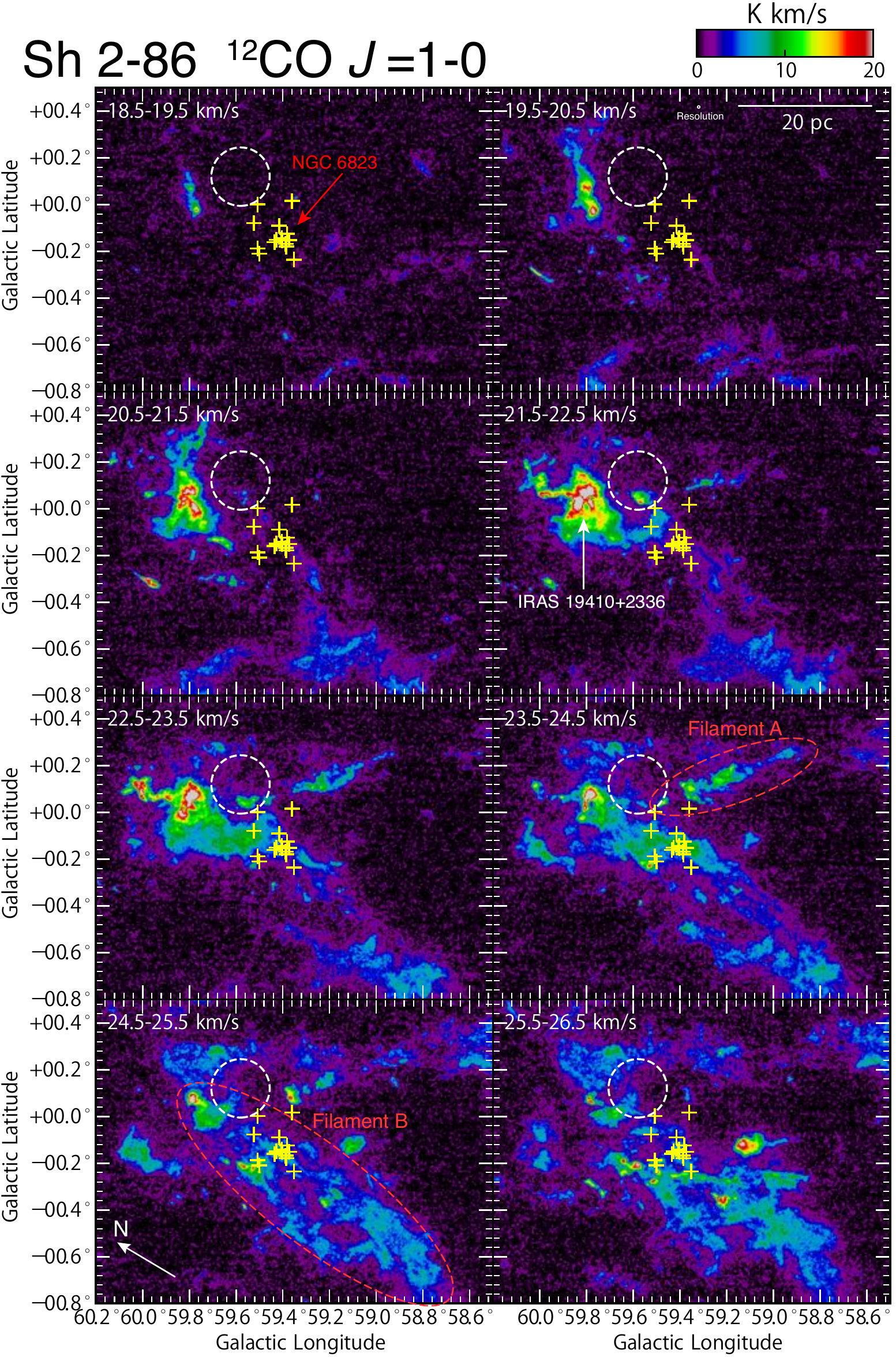}
\end{center}
\caption{Velocity-channel map of the $^{12}$CO $J=$ 1--0 emission with velocity steps of 1.0 $\>$km s$^{-1}$. {The plots are the same as in Figure\ref{Vul_HiGAL}(b). } {The map resolution} after convolution is \timeform{40"}. The $1\sigma$ noise level is $\sim 0.7$ K km s$^{-1}$ for the velocity interval of 1.0 km s$^{-1}$.}
\label{12COch}
\end{figure*}
\addtocounter{figure}{-1}

\begin{figure*}[h]
\begin{center} 
 \includegraphics[width=14cm]{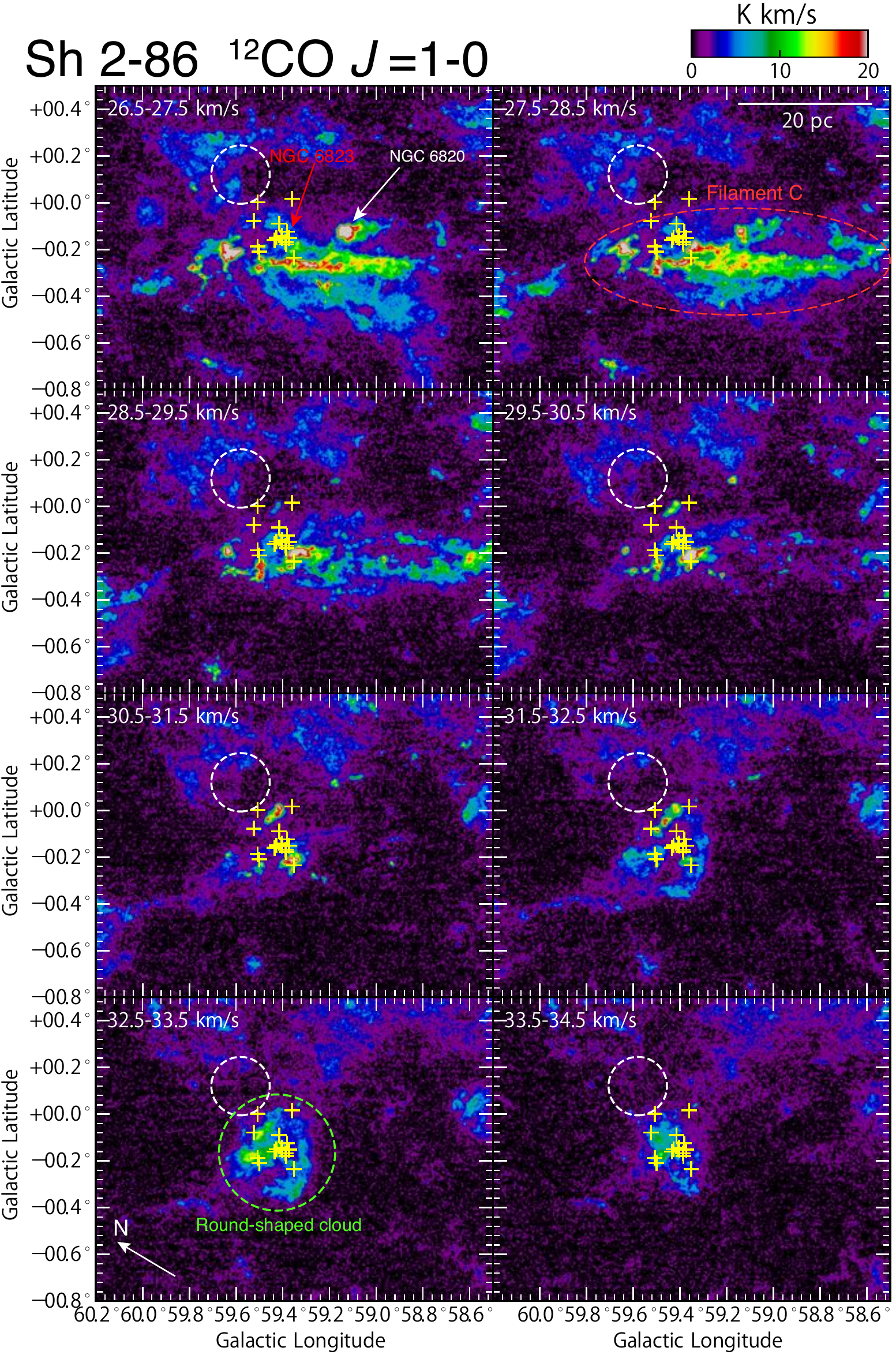}
\end{center}
\caption{{(Continued.)}}
\label{12COch2}
\end{figure*}

 \clearpage
\subsection{CO velocity distributions and the three velocity components in Sh 2-86}
\begin{figure*}[h]
\begin{center} 
 \includegraphics[width=14cm]{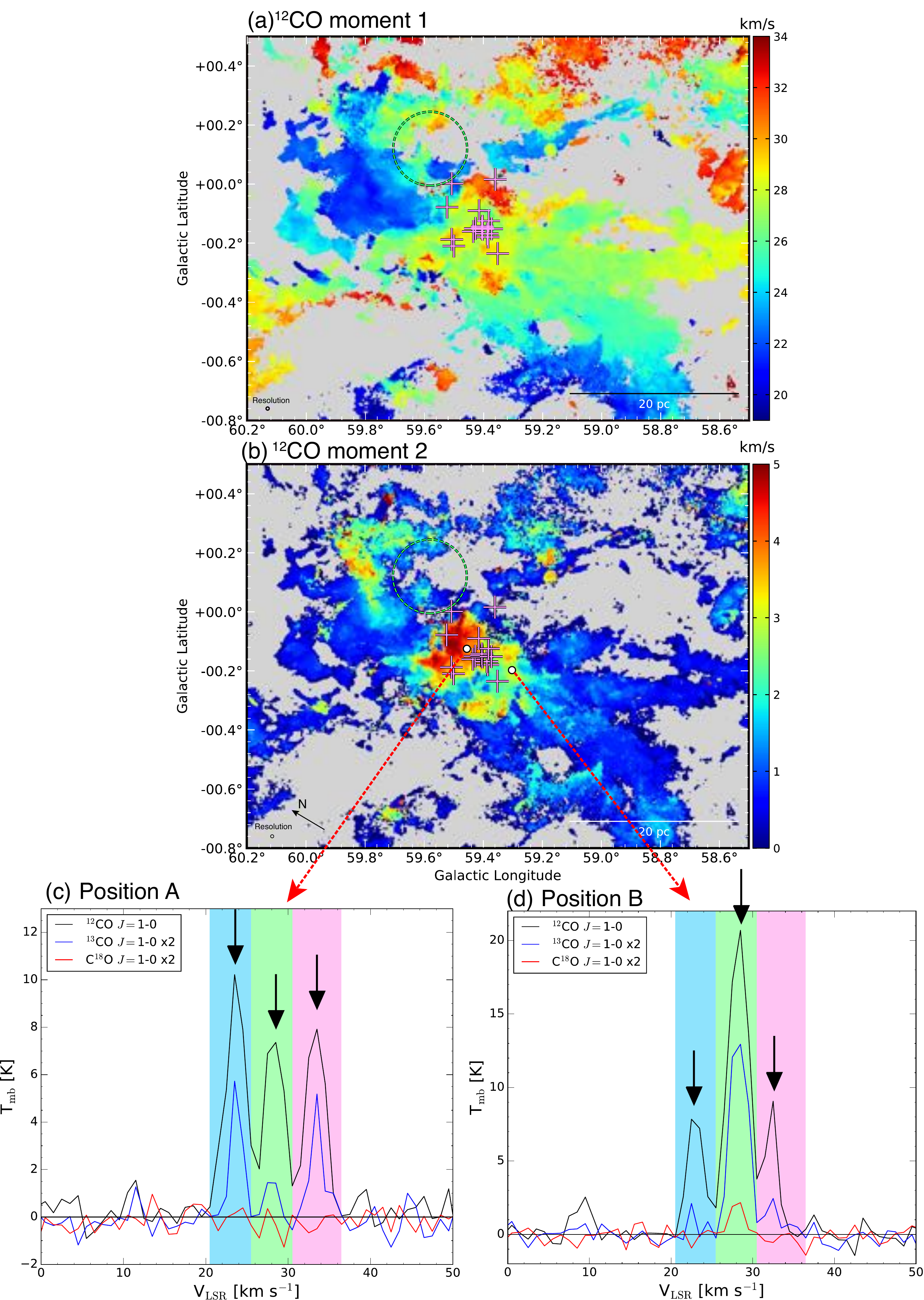}
\end{center}
\caption{(a) The $^{12}$CO first-moment (velocity-field) map {for} Sh 2-86. (b) The $^{12}$CO second-moment (velocity dispersion) map. The adopted velocity {range extends} from 19 to 34 km s$^{-1}$. {The crosses and dotted circles are the same as in Figure\ref{Vul_HiGAL}(b).} 
{Pixels are blanked if the line intensity is less than 3.5 K ($\sim 5 \sigma$).} {The map resolution} after convolution is \timeform{40"}. {(c) and (d) The CO spectra observed at two positions indicated by the white circles in panel. Short black arrows show the three velocity components at 22, 27, and 33 km s$^{-1}$. Blue, green, and red shadows show the integrated velocity range of each velocity component in this paper. }}
\label{12COmom}
\end{figure*}
Figure \ref{12COmom} {presents} the $^{12}$CO (a) intensity-weighted velocity map (first-moment) and (b) velocity dispersion (second-moment) maps. 
{We calculated the intensity-weighted velocity ($V_c$) and velocity dispersion ($\Delta V$) given by
\begin{eqnarray}
V_c &=& {\int T_{\rm B}(v)  v\ dv \over \int T_{\rm B}(v)\ dv} {\     [\rm km\ s^{-1}]},
\label{eq:vc}\\
\Delta V &=& \left\{{\int T_{\rm B}(v)  (v-V_c)^2 \ dv \over \int T_{\rm B}(v)\ dv}\right\}^{1/2} {\     [\rm km\ s^{-1}]}
\label{eq:dv}
\end{eqnarray}
,where $T_{\rm B}(v)$ and $v$ are the brightness temperature and radial velocity, respectively.}
The velocity field (Figure \ref{12COmom}a) {extends} $\sim 20$ km s$^{-1}$ around IRAS 19410+2336, and $\sim 28$ km s$^{-1}$ near the cluster {NGC 6823}.
 {A} large velocity dispersion of $\sim 5$ km s$^{-1}$ (Figure \ref{12COmom}b) exists at the open cluster NGC 6823, which also corresponds to the intersection of Filament B and Filament C.

{Figure \ref{12COmom} (c), (d) show the CO spectra near the NGC 6823 cluster at the position A and B.
We find the three velocity components at  22, 27, and 33 km s$^{-1}$. 
Blue, green, and red shadow area indicate the integrated velocity range of each component in this paper.}
We performed a detailed analysis of {these} multiple velocity components {with the goal of determining} the relationship 
between these molecular filaments {and} cluster formation in Sh 2-86.

\subsection{{$^{12}$CO $J=$2-1/1-0  intensity ratio and comparison to the {{Herschel}} dust-continuum images}}
\begin{figure*}[h]
\begin{center} 
 \includegraphics[width=17cm]{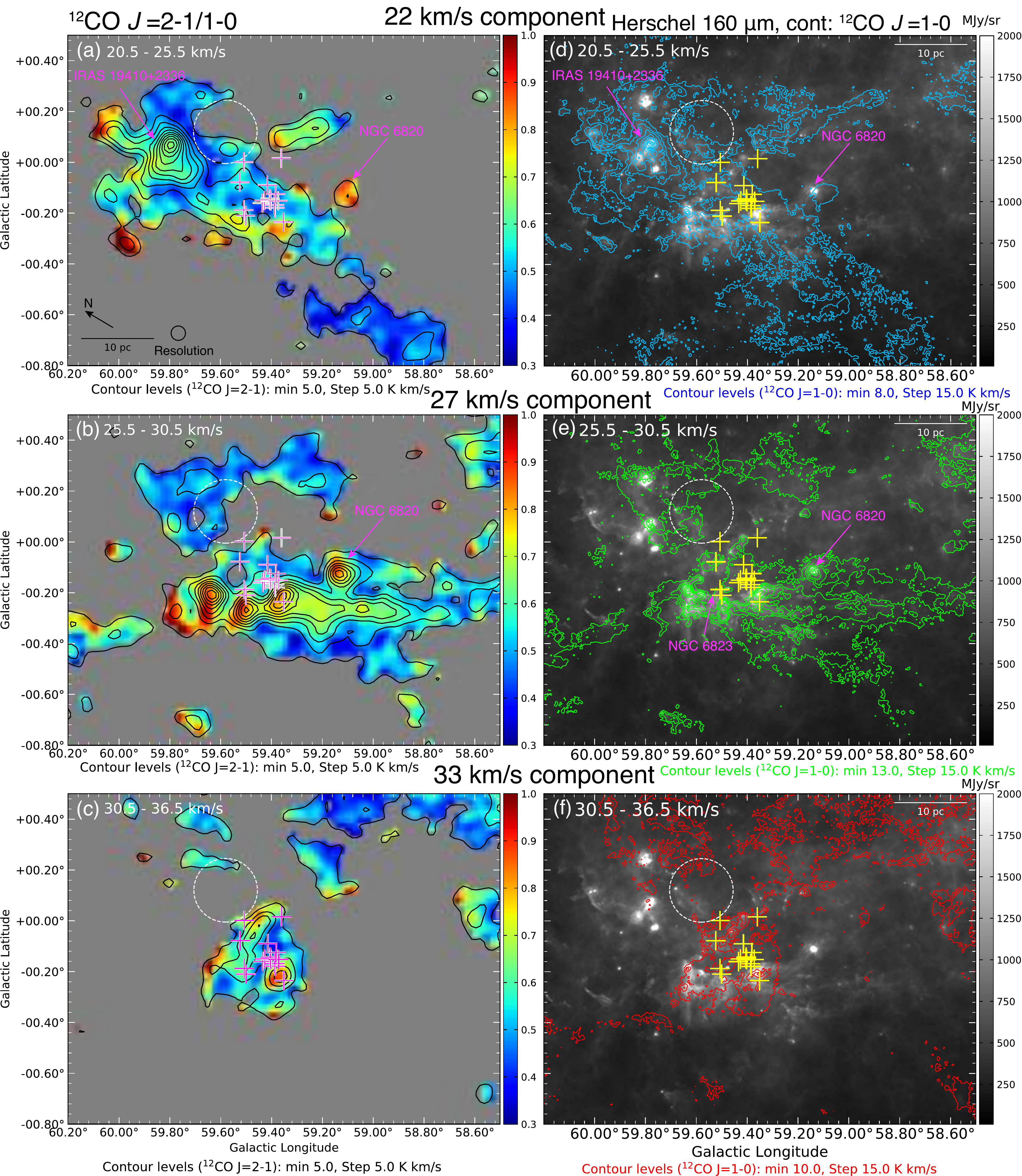}
\end{center}
\caption{{The $^{12}$CO $J=$2-1/1-0 intensity ratios of the (a) 22 km s$^{-1}$, (b) 27 km s$^{-1}$, and (c) 33 km s$^{-1}$ {components} in Sh 2-86. {The map resolution} after convolution is \timeform{3.4'}, which corresponds to the effective resolution of the $^{12}$CO $J=$2--1 data. {Pixels are blanked if the integrated intensity is less than 5 K km s$^{-1}$ ($\sim 5 \sigma$) for each integrated velocity range.} {The lowest contour levels and the contour intervals are 5.0 K km s$^{-1}$.} The integrated intensities (contours) of the three {components} are superposed on the {{Herschel}} 160 $\mu$m continuum image \citep{2016A&A...591A.149M} for the (d) 22 km s$^{-1}$, (e) 27 km s$^{-1}$, and (f) 33 km s$^{-1}$ clouds. 
{The lowest contour levels and the contour intervals are 8.0 K km s$^{-1}$ and 15.0 K km s$^{-1}$ for (d), 13.0 K km s$^{-1}$ and 15.0 K km s$^{-1}$ for (e), 10.0 K km s$^{-1}$ and 15.0 K km s$^{-1}$ for (f).} {The crosses and dotted circles are the same as in Figure\ref{Vul_HiGAL}(b)}}}
\label{Sh2-86_filament}
\end{figure*}


We performed detailed analyses of the {$^{12}$CO $J=$ 2-1/1-0} intensity ratio (hereafter $R_{\rm 2-1/1-0}$) and {compared it} with the infrared image toward these {velocity components}. 
Figures \ref{Sh2-86_filament}(a), (b), and (c) show the $R_{\rm 2-1/1-0}$ maps of {the} 22, 27, and 33 km s$^{-1}$ {components}, respectively.
The $^{12}$CO $J=$ 1-0 data {were} smoothed to \timeform{3.4'}, which corresponds to the effective resolution of the $^{12}$CO $J=$ 2-1 data. 
{Pixels are blanked if the integrated intensity is less than 5 K km s$^{-1}$ ($\sim 5 \sigma$) for each integrated velocity range.}
The intensity ratio of {the} different {CO} rotational levels is useful {for investigating} the excitation {states} of {the} molecular clouds.
The 22 km s$^{-1}$ {components} {has} a high-intensity ratio ($R_{\rm 2-1/1-0} \sim 0.8-1.0$) at NGC 6820 and {on the} eastern side of IRAS 19410+2336.
The 27 km s$^{-1}$ {components} has a high ratio ($R_{\rm 2-1/1-0} \sim 0.9-1.0$) at NGC 6823 and NGC 6820.
The 33 km s$^{-1}$ {components} has $R_{\rm 2-1/1-0} \sim 0.9-1.0$ around NGC 6823.
{These values are higher than the typical value of $R_{\rm 2-1/1-0} \sim 0.6$ in the Galactic plane (e.g., \cite{2010PASJ...62.1277Y}) and {in} external spiral galaxies (e.g., \cite{2020arXiv201208523Y}).
{The value of} $R_{\rm 2-1/1-0}$ depends on the kinematic temperature of {the molecular gas} and the number density of hydrogen {molecules}, assuming the large-velocity-gradient model (e.g., \cite{1974ApJ...189..441G}). Therefore, {a} high-intensity ratio {indicates} the presence of dense or warm gas associated with massive stars in {these} molecular clouds.
Figures \ref{Sh2-86_filament}(d), (e), and (f) {show the} spatial distributions of {the} three clouds (contours) superposed on the {{Herschel}} 160 $\mu$m image.
We find that {the} CO peaks in each velocity component {correspond morphologically} to the infrared peaks of IRAS 19410+2336, NGC 6823, and NGC 6820. 
From these results, we suggest that {these} three {components}{, which have} the velocity {separations} of 5-6 km s$^{-1}$ are likely to be physically associated with Sh 2-86.}

 \clearpage
\subsection{Physical parameters of {the} molecular gas}
\label{sec:phys}

\begin{table*}[h]
{
\tbl{Excitation temperature, optical depth, and column density of molecular clouds in Vul OB1.}{
{
\scalebox{1.0}{
\begin{tabular}{ccccccccccccccccc}
\hline
\multicolumn{1}{c}{Name} &$T_{\rm ex}$ & $\tau_{13}$ & $\tau_{18}$ &$N^{12}_{\rm X\ max}$ & $N^{12}_{\rm X\ mean}$ & $N^{13}_{\rm LTE\ max}$ & $N^{13}_{\rm LTE\ mean}$& $N^{18}_{\rm LTE\ max}$ & $N^{18}_{\rm LTE\ mean}$\\
&[K] & & & [cm$^{-2}$] & [cm$^{-2}$] & [cm$^{-2}$] & [cm$^{-2}$] & [cm$^{-2}$] & [cm$^{-2}$]  \\
(1) & (2) & (3) &(4)& (5) & (6) & (7) & (8) & (9) & (10)  \\
\hline\hline
Sh2-86  & 10 & 0.38  & 0.13 &$4.4 \times 10^{22}$&$8.0 \times 10^{21}$& $1.1 \times 10^{23}$ &$1.1 \times 10^{22}$ & $7.4 \times 10^{22}$& $1.6 \times 10^{22}$ \\
22 km s$^{-1}$ cloud & 10 & 0.38 & 0.13 & $2.9 \times 10^{22}$&$5.3 \times 10^{21}$& $1.0 \times 10^{23}$ &$1.1 \times 10^{22}$ & $7.4 \times 10^{22}$& $1.7 \times 10^{22}$ \\
27 km s$^{-1}$ cloud  & 10 & 0.38  & 0.1 &$2.2 \times 10^{22}$&$4.6 \times 10^{21}$& $7.0 \times 10^{22}$ &$7.7 \times 10^{21}$ & $5.3 \times 10^{22}$& $1.6 \times 10^{22}$ \\
32 km s$^{-1}$ cloud  & 8.8 & 0.41 &0.19 &$1.2 \times 10^{22}$&$2.8 \times 10^{21}$& $1.8 \times 10^{22}$ &$2.4 \times 10^{21}$ & $1.7 \times 10^{22}$& $1.0 \times 10^{22}$ \\
\hline
Sh2-87 &18 & 0.27 &0.04 &$6.0 \times 10^{22}$&$1.6 \times 10^{22}$& $2.0 \times 10^{23}$ &$3.7 \times 10^{22}$ & $8.0 \times 10^{22}$& $3.4 \times 10^{22}$ \\
Sh2-88  & 13 &0.39 & 0.09 &$6.4 \times 10^{22}$&$1.2 \times 10^{22}$& $2.7 \times 10^{23}$ &$2.7 \times 10^{22}$ & $1.5 \times 10^{23}$& $2.6 \times 10^{22}$ \\
\hline
\hline
\end{tabular}
}}}
\label{column_sh2-86}
\begin{tabnote}
Note: The mean values were adopted within the cloud surface area ($S$).
Columns: (1) Name. (2) The mean excitation temperature obtained from the $^{12}$CO peak intensity. (3) The mean optical depth obtained from $^{13}$CO. (4) The same as (3), but for C$^{18}$O. (5) The peak H$_2$ column density derived from $^{12}$CO assuming the X-factor. (6) The mean H$_2$ column density calculated from $^{12}$CO assuming the X-factor. (7) The same as (6), but for $^{13}$CO assuming LTE. (8) The same as (6), but for $^{13}$CO, assuming LTE. (9) The same as (7), but for C$^{18}$O (10) The same as (8), but for C$^{18}$O. 
\end{tabnote}
}
\end{table*}

\begin{table*}[h]
{
\tbl{Size and molecular mass in Vul OB1.}{
{
\scalebox{1.0}{
\begin{tabular}{ccccccccccccccccc}
\hline
\multicolumn{1}{c}{Name} &{$D_{12}$}& $M^{12}_{\rm X}$ &{$D_{13}$}& $M_{\rm LTE}^{13}$ & {$D_{18}$}&$M_{\rm LTE}^{18}$\\
&  [pc] &[$M_{\odot}$]& [pc]  &  [$M_{\odot}$]& [pc] &  [$M_{\odot}$]   \\
(1) & (2) & (3) &(4)& (5) & (6) & (7)  \\
\hline\hline
Sh2-86  & 43&$8.0 \times 10^4$ &15 & $4.3 \times 10^4$& 3.5 & $3.4 \times10^3$\\
22 km s$^{-1}$ component & 31 &$2.9 \times 10^4$&10 & $1.7 \times 10^4$& 3 & $1.8 \times10^3$\\
27 km s$^{-1}$ component & 34 &$4.5 \times 10^4$& 13 & $2.3 \times 10^4$& 2 & $1.3 \times10^3$\\
32 km s$^{-1}$ component & 22 &$1.4 \times 10^4$ & 10 &$4.1 \times 10^3$ & 0.7 &$78$\\
\hline
Sh2-87 & 7.4 &$6.5 \times 10^3$ & 3.6 &$8.4 \times 10^3$ &1.1 &$7.8 \times10^2$\\
Sh2-88 &15 &$1.1 \times 10^4$ & 4.9 &$1.1 \times 10^4$ &1.6 &$1.2 \times10^3$\\
\hline
\hline
\end{tabular}
}}}
\label{mass_sh2-86}
\begin{tabnote}
Note:{The physical parameters were calculated using the data points above $5 \sigma$ ($\sim 3.5$ K for $^{12}$CO and 0.90 K for $^{13}$CO and C$^{18}$O $J =$1--0).}
Columns: (1) Name. (2) {The $^{12}$CO cloud diameter within 10\% of the peak integrated intensity.}  (3) The total molecular mass estimated from $^{12}$CO. (4) The same as (2), but for $^{13}$CO. (5) The same as (3), but for $^{13}$CO. (6) The same as (2), but for C$^{18}$O. (7) The same as (3), but for C$^{18}$O. 
\end{tabnote}
}
\end{table*}

We derived the physical parameters of {the} molecular clouds in Vul OB1 following {a} procedure {that uses the} X-factor and {assumes} local thermal equilibrium {(LTE), which is} described by \citet{2008ApJ...679..481P,2013tra..book.....W}, and \citet{2020MNRAS}.

The brightness temperature $T_{\rm B}{(v)}$ {can be} expressed in {terms of the} excitation temperature $T_{\rm ex}$ and {the} optical depth $\tau {(v)}$, and {it is} given by

\begin{eqnarray}
T_{\rm B} {(v)} = T_0 \left(\frac{1}{\exp({T_0/ T_{\rm ex}})-1} - \frac{1}{\exp({T_0/ T_{\rm bg}})-1 }\right) \left(1-\exp({-{\tau (v)}})\right),
\label{Tb}
\end{eqnarray}
where $T_0$ is the Planck temperature defined {as $T_0 = {h\nu \over k}$, with $k$ and $h$ being the} Boltzmann and Plank constants, respectively, and  $T_{\rm bg}$ is {the} black-body temperature {of} the cosmic microwave background: $T_{\rm bg}=2.725 $ K.

If we assume that the $^{12}$CO $J=$1--0 line is {optically} thick ($\tau \to \infty$), the excitation temperature is given by

\begin{eqnarray}
T_{\rm ex} &=& 5.53 \bigg/ \ln \left(1+ {5.53  \over T_{\rm B}(\rm ^{12}CO)_{\rm max} /{\rm K} + 0.836 }\right) {\  [\rm K]},
\label{eq:tex}
\end{eqnarray}
where $T_{\rm B}(\rm ^{12}CO)_{\rm max}$ is the peak brightness temperature. {This yields} mean values $T_{\rm ex}\sim 10$ -- $18$ K in Vul OB1 and {in} each H\,\emissiontype{II} region.


The $^{13}$CO and C$^{18}$O optical {depths can be} derived from {the} brightness temperatures and the excitation temperature{, and they are given} by

\begin{eqnarray}
{\tau^{13}_0} &=& -\ln \left[1-{T_{\rm B}({\rm ^{13}CO})_{\rm max} \over 5.29\ {\rm K}} \left\{ {1 \over \exp({5.29\ {\rm K} \over T_{\rm ex}})-1}-0.168 \right\}^{-1} \right] {{\rm and}}
\label{eq:tau13}\\
{\tau^{18}_0} &=& -\ln \left[1-{T_{\rm B}({\rm C^{18}O})_{\rm max} \over 5.27\ {\rm K}} \left\{ {1 \over \exp({5.27\ {\rm K} \over T_{\rm ex}})-1}-0.169 \right\}^{-1} \right]{{\rm, respectively.}}
\label{eq:tau18}
\end{eqnarray}
The {values} ${\tau^{13}_0}$ and ${\tau^{18}_0}$ {at the peak intensity} {were thus found to be} $\sim 0.3$-$0.4$ and {$\sim 0.04$-$0.19$}, respectively, {for each H\,\emissiontype{II} region}.
In terms of the brightness temperature and {the} velocity width ($dv$), {the $^{13}$CO and C$^{18}$O column densities are} given by



\begin{eqnarray}
N ({\rm ^{13}CO}) &=& 2.42 \times 10^{14}  {T_{{\rm ex}}/ {\rm K}  \over 1-\exp \left(-{5.29\ {\rm K} \over T_{\rm ex} } \right) }  \int {\tau^{13} (v)}\  dv \nonumber \\
&{\sim} &2.42 \times 10^{14} {{\tau^{13}_0} \over 1-\exp(-{\tau^{13}_0})}  {1 \over 1-\exp \left(-{5.29\ {\rm K} \over T_{\rm ex} } \right) }  \int {T_{\rm B} (^{13}{\rm CO}) (v)}\  dv {\   [\rm cm^{-2}]}{,}
\label{eq:n13}\\
N ({\rm C^{18}O}) &=& 2.52 \times 10^{14} {T_{{\rm ex}}/ {\rm K}  \over 1-\exp \left(-{5.27\ {\rm K} \over T_{\rm ex} } \right) } \int  {\tau^{18} (v)}\ dv \nonumber \\
&{\sim}& 2.52 \times 10^{14} {{\tau^{18}_0} \over 1-\exp(-{\tau^{18}_0})} {1 \over 1-\exp \left(-{5.27\ {\rm K} \over T_{\rm ex} } \right) } \int {T_{\rm B} ({\rm C^{18}O}) (v)} \  dv  {\   [\rm cm^{-2}]}{,}
\label{eq:n18}
\end{eqnarray}
where {we use the relationship {$T_{\rm ex} \int \tau (v) dv \sim {\tau_0 \over 1-\exp(-\tau_0)} \int T_{\rm B} (v) dv$} {and assumes} the optically thin case quoted from \citet{2013tra..book.....W}.}

The H$_2$ column density is derived from the $^{12}$CO integrated intensity using the X-factor ($X({\rm ^{12}CO})$).
{The $^{13}$CO, and} C$^{18}$O column {densities are} calculated from the isotopic abundance {ratios using the following formulas}: 

\begin{eqnarray}
N^{12}_{\rm X} ({\rm H_2})  &=& X({\rm ^{12}CO}) \int T_{\rm B} (^{12}{\rm CO})\ dv,
\label{eq:x12}\\
N^{13}_{\rm LTE} ({\rm H_2}) &=& Y[{\rm ^{13}CO}] \times N ({\rm ^{13}CO}),
\label{eq:x13}\\
N^{18}_{\rm LTE} ({\rm H_2}) &=& Y[{\rm C^{18}O}] \times N ({\rm C^{18}O}).
\label{eq:x18}
\end{eqnarray}
We used $2.0\times 10^{20}$ [(K kms$^{-1}$)$^{-1}$ cm$^{-2}$] as the X-factor {whose uncertainty is 30\%.} \citep{2013ARA&A..51..207B}.
The {adopted} CO conversion {factors for} $Y[{\rm ^{13}CO}]$ and $Y[{\rm C^{18}O}]$ are $7.7 \times 10^5$ and $5.6 \times 10^6$, respectively.
These values are calculated from the isotopic abundance {ratios}  [$^{12}$C]/[$^{13}$C] $= 77$, {and} [$^{16}$O]/[$^{18}$O] $= 560$ \citep{1994ARA&A..32..191W}, and H$_2$ abundance ratio $[^{12}$CO]/[H$_2$] $= 10^{-4}$ (e.g., \cite{1982ApJ...262..590F,2010ApJ...721..686P}).
The peak column densities obtained {from} $^{13}$CO are $> 10^{23}$ cm$^{-2}$, which is larger than {that} estimated {from} $^{12}$CO.
This tendency is consistent with the other GMCs in the Milky Way (e.g., \cite{2020MNRAS}).

{We estimated the diameter $D$ of {each} molecular cloud using the equation
\begin{eqnarray}
D = 2 \sqrt{{S \over \pi}},
\label{dia}
\end{eqnarray}
where S is the cloud area {{within} 10\% of the peak integrated intensity}.
The {resulting} diameters of the clouds in Sh 2-86, Sh 2-87, and Sh 2-88 obtained by $^{12}$CO are $\sim 40$ pc, $\sim 7$ pc, and $\sim 15$ pc{, respectively}.}

The total molecular mass is {given} by

\begin{eqnarray}
M =  \mu_{\rm H_2} m_{\rm H} d^2 \sum_i {\Omega} N_i(\rm H_2),
\label{eq:mass}
\end{eqnarray}
where $\mu_{\rm H_2} \sim 2.8$ is the mean molecular weight of a hydrogen molecule including contribution of helium (e.g., Appendix A.1. of \cite{2008A&A...487..993K}), $m_{\rm H}=1.67 \times 10^{-24}$ g is 
the proton mass, $d$ is the distance to {the} Vul OB1 GMC, $\Omega$ is the solid angle {of the map pixel, and $N_i({\rm H_2})$ is the molecular column density of the i-th pixel.}
{The total mass of Sh 2-86, Sh 2-87, and Sh 2-88 are $8.0 \times 10^4, 6.5 \times 10^3$, and $1.1 \times 10^4\ M_{\odot}$ derived by $^{12}$CO. The mass derived from $^{12}$CO and $^{13}$CO is consistent {with this} within a factor of 2-3. 
On the other hand, the mass derived from C$^{18}$O is about smaller than one order of magnitude. 
This difference causes the C$^{18}$O line to be optically thin and trace the dense region in the Vul OB1 GMC.
These results are consistent with molecular clouds in the Galactic Plane revealed by the FUGIN CO survey (see Figure 13 of \cite{2019PASJ...71S...2T}}).
We summarize the physical parameters of molecular clouds in Vul OB1 in {Table \ref{column_sh2-86} and \ref{mass_sh2-86}}.





\clearpage
\section{Discussion}

\subsection{{Comparison of the H$_2$ column density with YSO distributions}}
\begin{figure*}[h]
\begin{center} 
 \includegraphics[width=14cm]{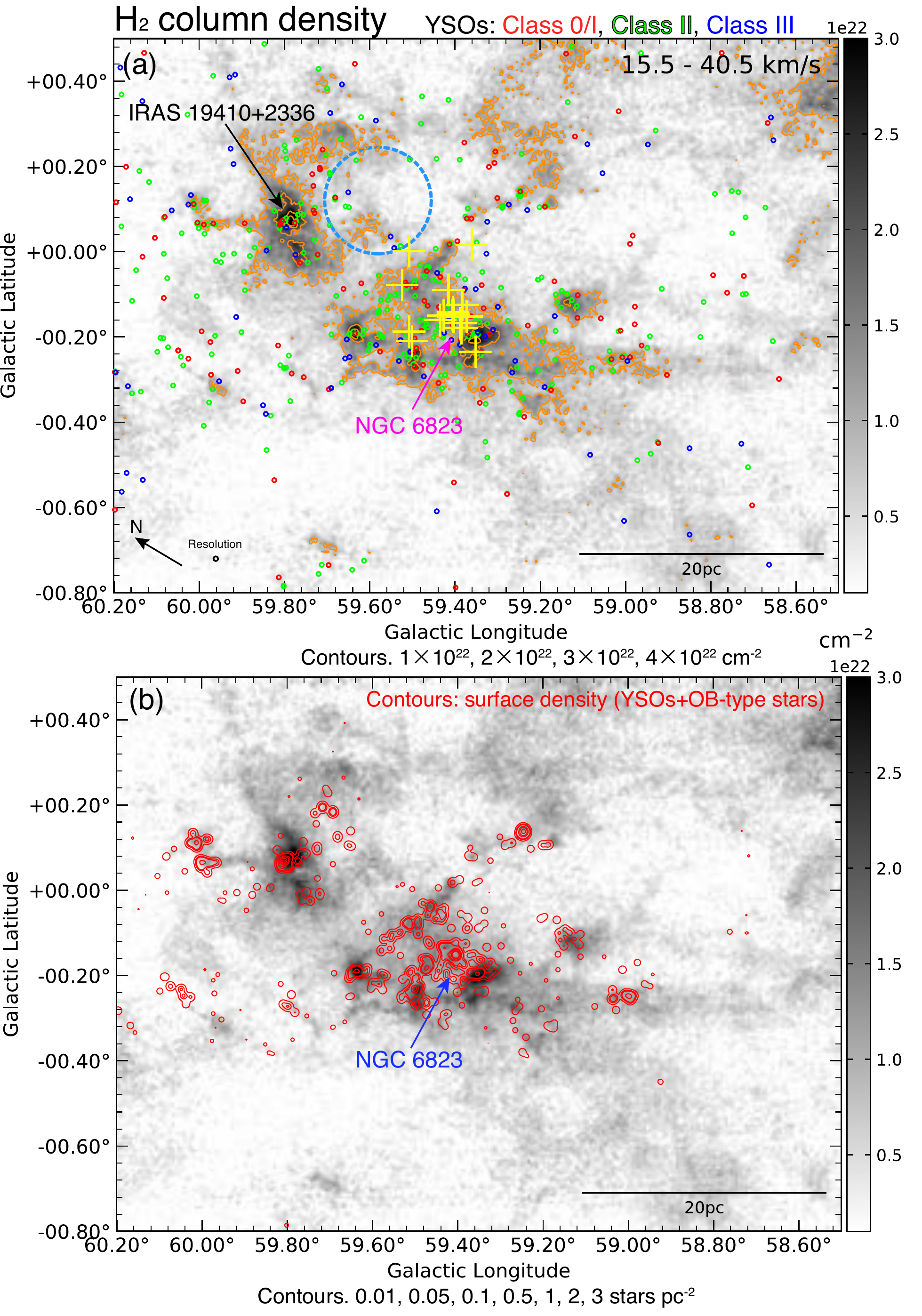}
\end{center}
\caption{{{(a)} The YSO distributions superposed on the H$_2$ column density maps derived from the $^{12}$CO $J=$1--0 data. 
{The contour levels for the column density are $1\times 10^{22}, 2\times 10^{22}, 3\times 10^{22}, {\rm and\ } 4\times 10^{22}$ cm$^{-2}$. } Red, green, and blue circles indicate the class 0/I, class II, and class III objects, respectively, as identified by \citet{2010ApJ...712..797B}. The yellow crosses indicate the OB-type stars in NGC 6823 identified by \citet{1995ApJ...454..151M}. The blue dotted circle shows the position of SNR G59.5+0.1 \citep{2019JApA...40...36G}. 
{(b) The surface density of YSOs and OB-type stars (red contours) superposed on the H$_2$ column density map. The red contours were convolved to \timeform{1'} with a Gaussian kernel function.} {The contour levels for the surface density are 0.01, 0.05, 0.1, 0.5, 1, 2, and 3 stars pc$^{-2}$. } The {column density map resolution} after convolution is \timeform{40"}.}}
\label{12CO_YSO}
\end{figure*}

Figure \ref{12CO_YSO}{(a)} shows the spatial distribution of {the} YSOs cataloged by \citet{2010ApJ...712..797B} superposed on the H$_2$ column density map derived from {the} $^{12}$CO $J=$ 1--0 integrated intensity using the X-factor $X({\rm CO}) = 2.0 \times 10^{20}$ [(K kms$^{-1}$)$^{-1}$ cm$^{-2}$] \citep{2013ARA&A..51..207B}.
The column density {is} high ($>1.0 \times 10^{22}$ cm$^{-2}$) near IRAS 19410+2336 and NGC 6823. 
{Figure \ref{12CO_YSO}(b) presents the surface density of YSOs and OB-type stars (red contours) superposed on the H$_2$ column density map. 
We calculated the surface density of stars at each point of a \timeform{10"} grid using the nearest neighbor method \citep{2008ApJ...682..445C,2010ApJ...712..797B}. The equation of the surface density ($\sigma_{\rm star}$) is given by 
\begin{eqnarray}
\sigma_{\rm star}  &=& {N \over \pi r^2_{\rm N}}\  [{\rm stars\ pc}^{-2}],
\label{eq:density}
\end{eqnarray}
where $r_{\rm N}$ is the distance to the nearest neighbor. In this paper, we adopted $N=5$, the same as \citet{2010ApJ...712..797B}. 
The surface density map was convolved to \timeform{1'} with a Gaussian kernel function.
We point out that {the} YSOs and OB-type stars {are} {concentrated} around the NGC 6823 cluster. }
{\citet{2010ApJ...712..797B} {argue} that each {YSO-population class is homogenously distributed} in Vul OB1, and {they were unable to determine} the age sequence.
Hence, {they} excluded the {propagating} star formation scenario proposed by \citet{2001A&A...374..682E}. 
{However, we} note that the YSOs around SNR G59.5+0.1 {may have been} induced by the stellar wind from the G59.5+0.1 progenitor, {as} suggested by \citet{2008ApJ...680..446X}.
{{Based on} our observational results and previous studies, we {discuss the cluster formation scenario in NGC 6823 in the following subsection.}
}

\clearpage
\subsection{{The interpretations of the three velocity components associated with Sh 2-86 }}

\begin{figure*}[h]
\begin{center} 
 \includegraphics[width=18cm]{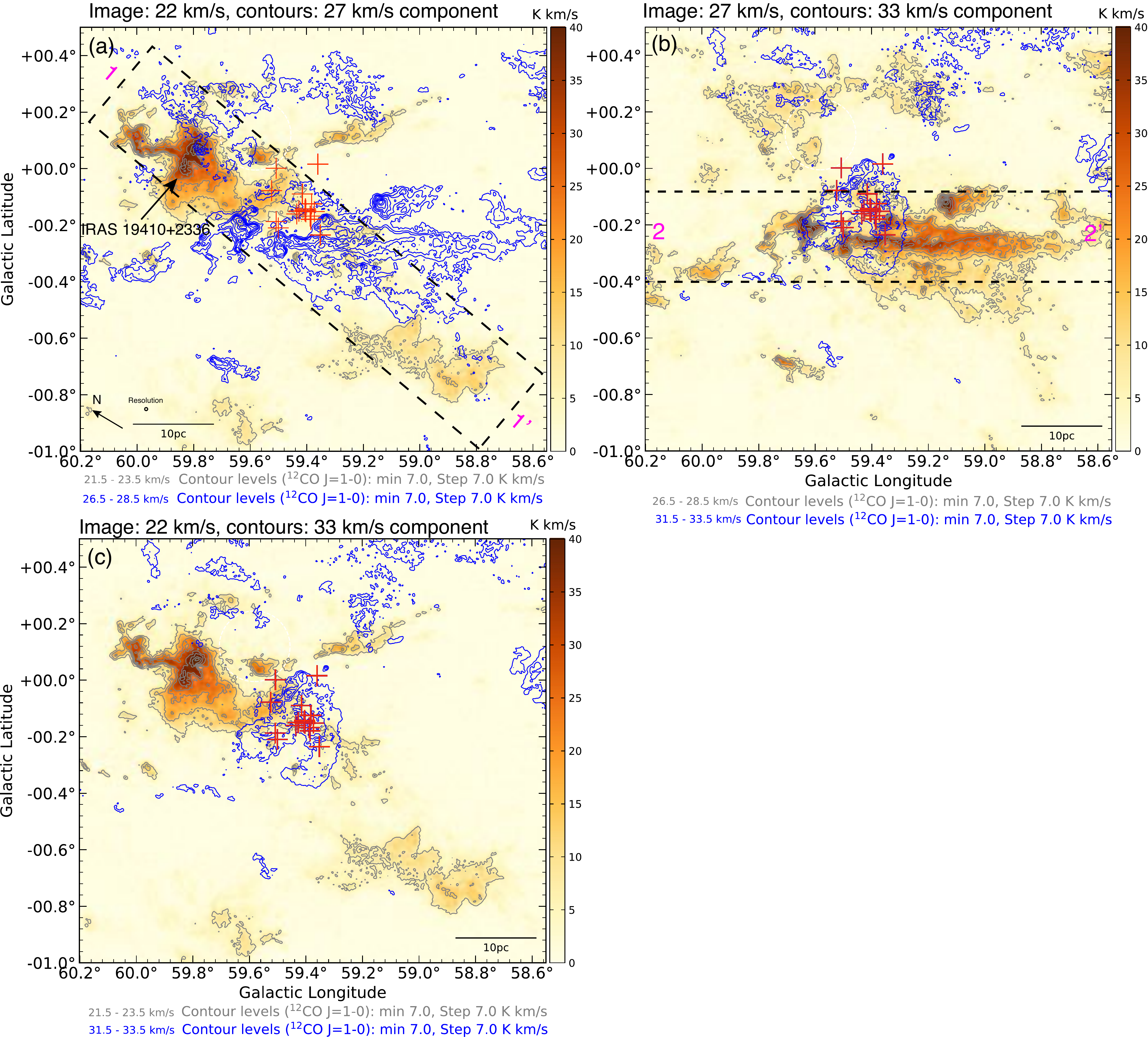}
\end{center}
\caption{{Spatial distributions of the $^{12}$CO $J=$1-0 emission integrated over two different velocity ranges.
 (a) The 22 km s$^{-1}$(image) and 27 km s$^{-1}$ (blue contours) {components}.}
 (b) The 27 km s$^{-1}$(image) and 33 km s$^{-1}$ (blue contours) {components}. 
  {Black dotted boxes shows the making ranges of the position-velocity diagram along the filaments in Figure \ref{Sh2-86_spec}(a) and (b).
 (c) The 22 km s$^{-1}$(image) and 33 km s$^{-1}$ (blue contours) {components}. 
{The lowest contour levels and the contour intervals are 7.0 K km s$^{-1}$.} {The crosses are the same as in Figure \ref{12CO_YSO}. The map resolution} after convolution is \timeform{40"}. }}
\label{12CO_overlay}
\end{figure*}

{Figure \ref{12CO_overlay} presents the $^{12}$CO $J=$1-0 integrated intensity maps {for the two} overlapping clouds associated with Sh 2-86.
The 27 km s$^{-1}$ cloud has peaks at {the} southern side of the cluster, corresponding to the intensity depression of the 22 km s$^{-1}$ cloud (Figure  \ref{12CO_overlay}a). {That shows anti-correlated spatial distributions.}
 The {NGC 6823} cluster members {are found to be} the overlap region between {the 33 km s$^{-1}$ cloud and 22, 27 km s$^{-1}$ components (Figure \ref{12CO_overlay}b, c)}. 
These components are also connected in {the} position-velocity diagram (see Figure \ref{Sh2-86_spec}).


\begin{figure*}[h]
\begin{center} 
 \includegraphics[width=18cm]{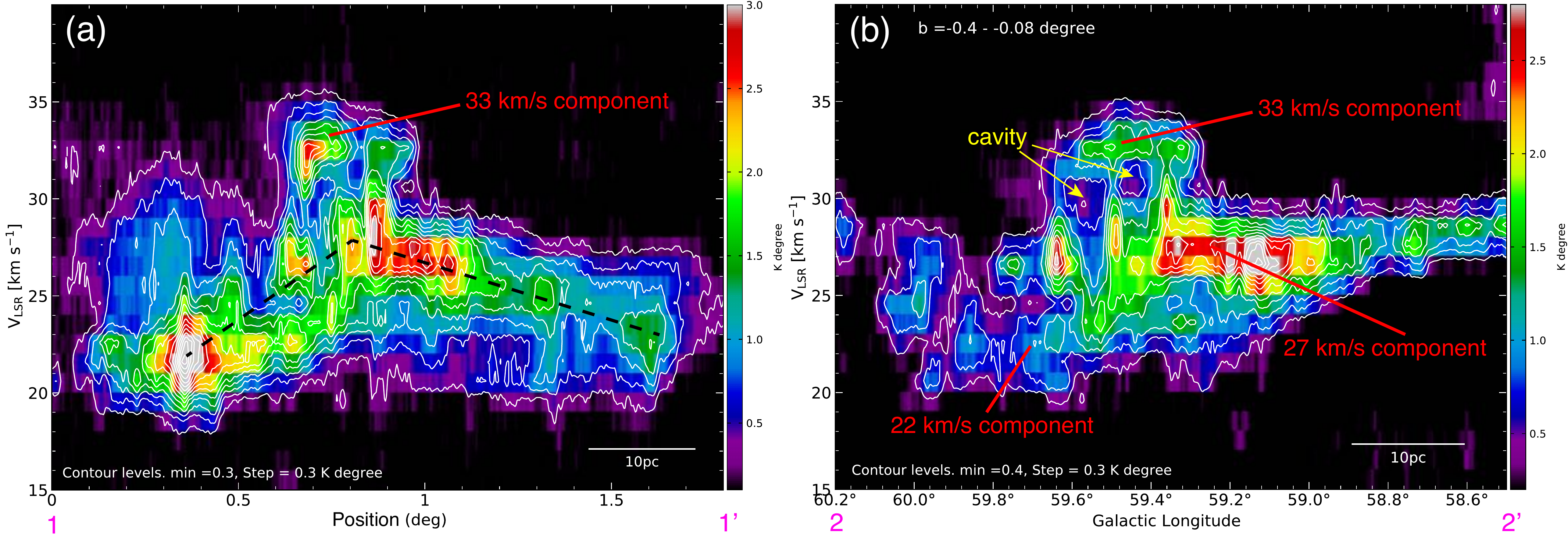}
\end{center}
\caption{{(a) Position-velocity diagram for $^{12}$CO integrated along the Filament B (cut 1-1' in Figure \ref{12CO_overlay}a). The {lowest contours} and intervals are 0.3 K degree and 0.3 K degree, respectively. (b) The same as (a), but along the Filament C (cut 2-2' in Figure \ref{12CO_overlay}b).The {lowest contours} and intervals are 0.4 K degree and 0.3 K degree, respectively. }}
\label{Sh2-86_spec}
\end{figure*}


{Molecular clouds associated with massive star-forming regions often compose of multiple velocity components. They are interpreted as the expansion by stellar wind (e.g., \cite{2010A&A...523A...6D}), oscillation of a single cloud (e.g., \cite{2003ApJ...586..286L}),  infall motion of a massive cloud (e.g., \cite{2016ApJ...832..205S,2018ApJ...855...45S}), and a cloud-cloud collision (e.g., Kohno et al. 2021a, 2021b). 
{To investigate the origin of three velocity components in detail}, we made the position-velocity diagram along with the elongation of Filament B and C as shown in Figure \ref{Sh2-86_spec}(a) and (b), respectively. 
}
{\citet{2010ApJ...712..797B} and \citet{2008ApJ...681..428C} pointed out that the ionization feedback by the stellar wind from the OB-type stars in the NGC 6823 cluster 
affects the parent molecular cloud of Sh 2-86.
The two cavity structures having $<$5 pc exist at $\sim 31$ km s$^{-1}$ presented in Figure \ref{Sh2-86_spec}(b). 
We suggest that they were formed by the ionization effect of the NGC 6823 cluster. 
On the other hand, the 22 km s$^{-1}$ component exists 5 km s$^{-1}$ apart from the cavity velocity and distributes at the blue-shifted side of the 27 km s$^{-1}$ component. 
{The 22 km s$^{-1}$ component also has an intensity peak at IRAS 19410+2336, which is located $\sim 15$ pc away from the NGC 6823 cluster (Figure \ref{12CO_overlay}a).}
Therefore, we suggest that the effect of stellar feedback by the NGC 6823 cluster toward the 22 km s$^{-1}$ component is limited.}



{\citet{2016ApJ...832..205S} reported the two velocity components having the velocity separation of $\sim 1 $ km s$^{-1}$ associated with the massive clump of S235 AB. 
{They showed symmetrical two peaks with the clump center at the position-velocity diagram taken along its major axis (see Figure 3b in \cite{2016ApJ...832..205S}).}
The authors argued that their velocity distribution is interpreted as the infall motion of a single massive cloud with rotation based on the model of single protostars presented by \citet{1997ApJ...475..211O} (see also Type 2 massive clumps presented by \cite{2018ApJ...855...45S}). 
{In the case of Sh 2-86, we find the inverted V-shaped structure (Figure \ref{Sh2-86_spec}a) and three intensity peaks (Figure \ref{Sh2-86_spec}b) on the position-velocity diagrams along with Filament B and Filament C, respectively. These features are different from the velocity gradients reported by \citet{2016ApJ...832..205S}. }
Hence, we suggest that the infall motion with rotation of the GMC scale is unlikely to explain our observational results.}

{\citet{2003ApJ...586..286L} proposed that the oscillation of a single cloud can explain the origin of the velocity gradient in the molecular cloud associated with the starless core Barnard 68. 
{If the cloud has oscillation, we are likely to find the velocity pattern like the sine wave on the position-velocity diagram along the filament (e.g., Figure 6 in \cite{2019MNRAS.487.1259L}). However, we cannot find them on the position-velocity diagram in Sh 2-86.}
Thus, we suggest that it is difficult for the cloud oscillation (pulsation) to explain the origin of the three independent velocity components associated with Sh 2-86.}

{Based on our CO data, we show the connecting the three velocity components in the position-velocity diagram, V-shape structure, complementary spatial distributions, and high-intensity ratio around the NGC 6823 cluster. 
We propose that these observational signatures show the evidences of cloud-cloud collisions (see also the review paper by \cite{Fukui+2020}). 
Numerical simulations of cloud-cloud collisions reproduced the (inverted) V-shaped structures (see Figure 5 middle panel in \cite{2015MNRAS.450...10H}), and anti-correlated spatial distributions of different velocity channels \citep{2015ApJ...801...77M}. 
\citet{2018PASJ...70...96A} also argue that V-shaped structure shows the shock compression for the filament formation by colliding flows. 
{Theoretical studies have also demonstrated} that cloud-cloud {collisions} can set up the initial conditions for massive cluster star formation (e.g., \cite{1992PASJ...44..203H,2013ApJ...774L..31I,2018PASJ...70S..53I,2014ApJ...792...63T,2015ApJ...811...56W}). 
The radio observations were also suggested that cloud-cloud collisions (e.g., Sagittarius B2: \cite{1994ApJ...429L..77H}, M20: \cite{2011ApJ...738...46T}, RCW 38: \cite{2016ApJ...820...26F}, IRAS 05358+3543: \cite{2021arXiv210601852Y}) and filament-filament collisions (e.g., Serpens South: \cite{2014ApJ...791L..23N}, Cygnus OB 7: \cite{2014ApJ...797...58D}) as a trigger of massive stars and cluster formation in the Milky Way. {These results were also supported by the recent observational studies of Large Magellanic Cloud (e.g., \cite{2015ApJ...807L...4F,2017ApJ...835..108S, 2019ApJ...886...14F,2019ApJ...886...15T}), M33 (e.g., \cite{2021PASJ...73S..62S}) and Antennae Galaxies (e.g., \cite{2021PASJ...73..417T,2021PASJ...73S..35T}).} }

{\citet{2008ApJ...680..446X} and \citet{2010ApJ...719.1813H} argue that the clump-clump collision triggered the cluster formation {of Sh 2-87 and Sh 2-88 in Vul OB1}. 
In the case of Sh 2-86, it is excited by the 17 OB-type stars in the NGC 6823 cluster (see Table 2). This H\,\emissiontype{II} region is also the most active star-forming site in Vul OB1\citep{2010ApJ...712..797B}.
We suggest that multiple collisions between molecular cloud and GMFs having high peak column density $\sim 1 \times 10^{23}$ cm$^{-2}$ obtained by $^{13}$CO are likely to explain the origin of the multiple OB-type star formation in Vul OB1. 
The dynamical {collision timescale} is estimated {to be} 20-30 pc/ (5 km s$^{-1}$ $\times \sqrt{2}$)$\sim $ 3-4 Myr, assuming {a} viewing angle of $\timeform{45D}$. This value is roughly consistent with the cluster age of NGC 6823 ($3\pm 1$ Myr: \cite{2000AcA....50..113P}).
Therefore, we propose multiple cloud collisions are the most plausible hypothesis to explain the origin of the three velocity components and OB-type stars in Sh 2-86. 
To advance our current understanding in Vul OB 1, additional molecular line observations using shock tracers, such as SiO \citep{2020PASJ..tmp..238F,2020MNRAS.499.1666C} and dense gas tracers \citep{2021MNRAS.506..775P}, are helpful, leading the dynamical state as a consequence of the molecular cloud collisions.}} 
{
\subsection{The origin of {the} filamentary clouds in the Local Spur of the Milky Way}


\begin{table*}[h]
{
\tbl{Comparison with physical parameters of GMFs as sites of a cloud-cloud collision}{
\begin{tabular}{ccccccccccccccc}
\hline
\multicolumn{1}{c}{Name}& Length & Width  & $v_{\rm FWHM}$ &$N({\rm H_2})_{\rm peak}$ &Total mass & $M_{\rm line}$ &Reference \\
&[pc] & [pc] & [km s$^{-1}$] & [cm$^{-2}$] &[$M_{\odot}$] & [$M_{\odot}/$pc] &\\
(1) & (2) & (3) &(4)& (5) & (6) & (7) & (8) \\
\hline
Filament C in Sh 2-86 & $\sim 30$ & $\sim 5$  & $\sim 3$ & $ 1 \times 10^{22}$ &$\sim 4 \times 10^4$ &$\sim 1 \times 10^3$& This work\\
HVS in W51$^*$ & $\sim 90$ & $\sim 10$  & $\sim 5$  & $ 2 \times 10^{22}$ & $\sim 3 \times 10^5$ & $\sim 3 \times 10^3$&\citet{2019PASJ..tmp...46F}\\
Filament B in GMC-16 $^{\dag}$ & $\sim 70$& $\sim 5$-$6$&  $\sim 4$ &$ 5 \times 10^{22}$& $\sim 4 \times 10^5$ & $\sim 6 \times 10^3$ &\citet{2020ApJ...896...36T}\\
\hline
\end{tabular}}
\label{comp_GMF}
\begin{tabnote}
$*$ The parameters of the W51 GMC was taken from the FUGIN CO survey data (\cite{2017PASJ...69...78U,2019PASJ..tmp...46F}; URL: \url{https://nro-fugin.github.io}).\\ 
${\dag}$ The parameters of Filament B in GMC-16 of M33 were obtained by the $^{12}$CO $J=$2--1 data. We applied the X(CO) factor of $4 \times 10^{20}$ [(K kms$^{-1}$)$^{-1}$ cm$^{-2}$] and the $^{12}$CO $J=$2--1/1--0 intensity ratio of 0.7.\citep{2010A&A...522A...3G,2014A&A...567A.118D}. More detailed information is described in \citet{2020ApJ...896...36T} and \citet{2021arXiv210301610K}. \\
Columns: (1) Name. (2) The GMF length.  (3) The GMF width. (4) The FWHM ($2 \sigma \sqrt{2 \ln 2}$) of peak spectra by a single Gaussian fitting, where $\sigma$ is the standard deviation of the line profile. (5) The H$_2$ peak column density assuming the X-factor. (6) The total molecular mass derived by $^{12}$CO. (7) Line mass obtained by the total mass divided GMF length. (8) Reference.
\end{tabnote}
}
\end{table*}

In particular, Filament C in the 27 km s$^{-1}$ cloud {is distributed} along the Galactic plane.
This feature is common {to} the high-velocity stream (HVS) in the W51 GMC \citep{2019PASJ..tmp...46F}. 
{A} galactic spiral density wave {has been suggested} as the origin of the HVS (e.g \cite{1970A&AS....2..291B,1998AJ....116.1856C,1999ApJ...518..760K,2004MNRAS.353.1025K}).
Recently, \citet{2020ApJ...896...36T} discovered the GMFs associated with the high-mass star-forming region in {the} external spiral galaxy M33. The authors suggested that a Galactic spiral shock {had formed these} GMFs.   
{ Table \ref{comp_GMF} shows the comparison of Filament C with the physical parameters of GMFs {as sites of a cloud-cloud collision. }
Filament C has $\sim$ 1/2-1/3 length comparing with HVS and GMC-16. 
This value is also smaller than other GMFs identified by previous studies with $\sim 100$ pc in the Galactic plane (e.g., \cite{2010ApJ...719L.185J,2014A&A...568A..73R}).
The total mass is $\sim$ 1/10 of HVS and GMC-16. 
Therefore, we suggest that Filament C at the Local spur of the Milky Way is the smaller scale GMF than reported by previous studies. }

On the other hand, {based on the observations} of the external spiral galaxy M51 (e.g., \cite{2009ApJ...700L.132K, 2014PASJ...66...36M}), 
 it has been suggested that filamentary clouds in inter-arm regions can be produced by galactic shear motions.
 {Numerical simulations by \citet{2004MNRAS.349..270W} and \citet{2006MNRAS.367..873D} 
 have reproduced such filamentary structures in inter-arm regions.}
Hence, we suggest that they are formed by galactic-scale dynamics like {spiral shocks} or {shear motions}.
More detailed analyses are necessary to {confirm} the origins of the GMFs in Sh 2-86.
{{With such} detailed {considerations}, we will argue the separated paper.}

\section{Summary}
The conclusions of this paper are as follows:
{
\begin{enumerate}
\item We have performed large-scale $^{12}$CO, $^{13}$CO, {and} C$^{18}$O $J=$1--0 observations toward the Vulpecula OB association ($l \sim \timeform{60D}$) as a part of the Nobeyama 45 m Local Spur CO survey project, with the goal of determining  the cause of cluster formation in an inter-arm {region} of the Milky Way. 
\item {The molecular} clouds are distributed over {$\sim 100$} pc, and {they} have peaks at {the HII regions} Sh 2-86, Sh 2-87, and Sh 2-88 in Vul OB1.  
The whole of Vul OB1 {is located} in the Local Spur between the Local {Arm} and the Sagittarius {Arm} of the Milky Way.

\item {We discovered new {GMFs} in Sh 2-86, which have a {length} of $\sim 30$  pc, {width} of $\sim 5$ pc, and {mass} of $\sim 4\times 10^4\ M_{\odot}$ {obtained by $^{12}$CO}}. 

\item Sh 2-86 {contains} three velocity components {at} 22, 27, {and} 33 km s$^{-1}$. These clouds have high intensity-ratios around high-mass stars, and 
{they {coincide with}} to the infrared dust emission. Therefore, we conclude that {the} multiple velocity components are likely to be physically associated with Sh 2-86.

\item The OB-type stars in the open cluster NGC 6823 {are located} at the intersection of three clouds. From these observational results, we suggest that 
the cloud-cloud collision scenario can explain the origin of cluster formation in the Vul OB1 GMC.

\item {We propose that {the} GMFs in Vul OB1 {were} produced by galactic-scale dynamics like {spiral shocks or shear motions}.}

\end{enumerate}
}

\section*{Acknowledgements}
{The authors are grateful to the referee for thoughtful comments on the paper.
We are grateful to Mr. Rin Yamada of Nagoya University for a useful discussion.}

The Nobeyama 45-m radio telescope is operated by Nobeyama Radio Observatory, a branch of the National Astronomical Observatory of Japan.
{{Herschel}} is an ESA space observatory with science instruments provided by European-led Principal Investigator consortia and with important participation from NASA.
PACS has been developed by a consortium of institutes led by MPE (Germany) and including UVIE (Austria); KU Leuven, CSL, IMEC (Belgium); CEA, LAM (France); MPIA (Germany); INAF-IFSI/OAA/OAP/OAT, LENS, SISSA (Italy); IAC (Spain). 
This development has been supported by the funding agencies BMVIT (Austria), ESA-PRODEX (Belgium), CEA/CNES (France), DLR (Germany), ASI/INAF (Italy), and CICYT/MCYT (Spain).
"SPIRE has been developed by a consortium of institutes led by Cardiff University (UK) and including Univ. Lethbridge (Canada); NAOC (China); CEA, LAM (France); IFSI, Univ. Padua (Italy); IAC (Spain); Stockholm Observatory (Sweden); Imperial College London, RAL, UCL-MSSL, UKATC, Univ. Sussex (UK); and Caltech, JPL, NHSC, Univ. Colorado (USA). This development has been supported by national funding agencies: CSA (Canada); NAOC (China); CEA, CNES, CNRS (France); ASI (Italy); MCINN (Spain); SNSB (Sweden); STFC, UKSA (UK); and NASA (USA)."

The work is financially supported by a Grant-in-Aid for Scientific Research (KAKENHI, No. 18K13580, ) from MEXT (the Ministry of Education, Culture, Sports, Science and Technology of Japan) and JSPS (Japan Society for the Promotion of Science).

The authors would like to thank Enago (www.enago.jp) for the English language review.

Software: We utilized Astropy, a community-developed core Python package for astronomy \citep{2013A&A...558A..33A, 2018AJ....156..123A}, NumPy \citep{2011CSE....13b..22V}, Matplotlib \citep{2007CSE.....9...90H}, IPython \citep{2007CSE.....9c..21P}, APLpy \citep{2012ascl.soft08017R}, Miriad \citep{1995ASPC...77..433S}, and Montage \footnote{http://montage.ipac.caltech.edu} software {\citep{2017PASP..129e8006B}}.

{This research made use of Montage. It is funded by the National Science Foundation under Grant Number ACI-1440620, and was previously funded by the National Aeronautics and Space Administration's Earth Science Technology Office, Computation Technologies Project, under Cooperative Agreement Number NCC5-626 between NASA and the California Institute of Technology.}

\appendix
\section{{$^{13}$CO and  C$^{18}$O velocity channel maps of Sh 2-86}}
We present the velocity channel maps of the $^{13}$CO and C$^{18}$O $J =$ 1-0 emissions toward Sh 2-86 in the Appendix 1 (see Figures {\ref{13COch1} and \ref{C18Och1}})

\begin{figure*}[h]
\begin{center} 
 \includegraphics[width=14cm]{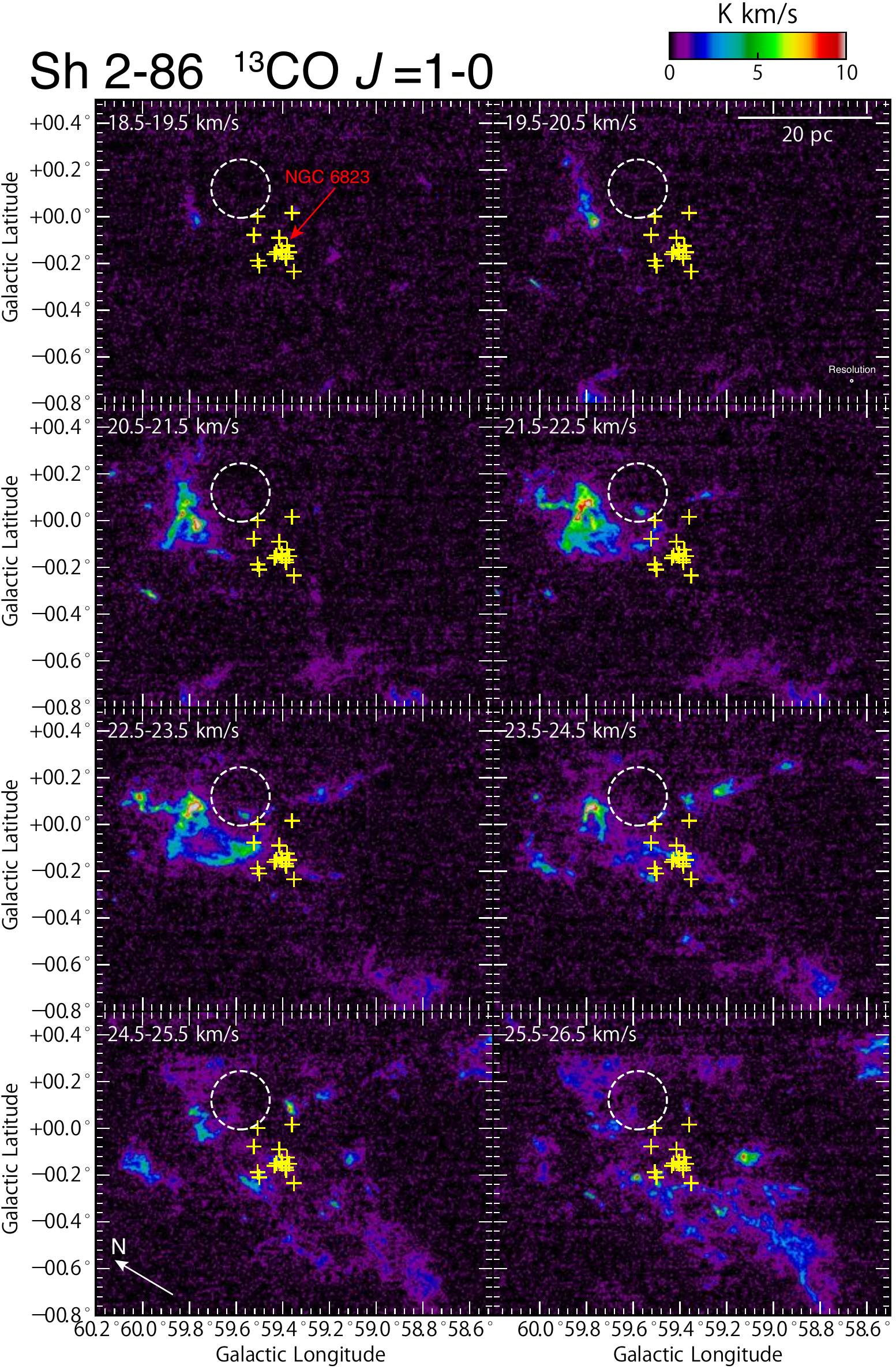}
\end{center}
\caption{Velocity-channel map of the $^{13}$CO $J=$ 1--0 emission with a velocity step of 1.0 $\>$km s$^{-1}$. {The crosses and dotted circles are the same as in Figure\ref{Vul_HiGAL}(b). } {The map resolution} after convolution is \timeform{40"}. The $1\sigma$ noise level is $\sim 0.3$ K km s$^{-1}$ for the velocity interval of 1.0 km s$^{-1}$.}
\label{13COch1}
\end{figure*}
\addtocounter{figure}{-1}

\begin{figure*}[h]
\begin{center} 
 \includegraphics[width=14cm]{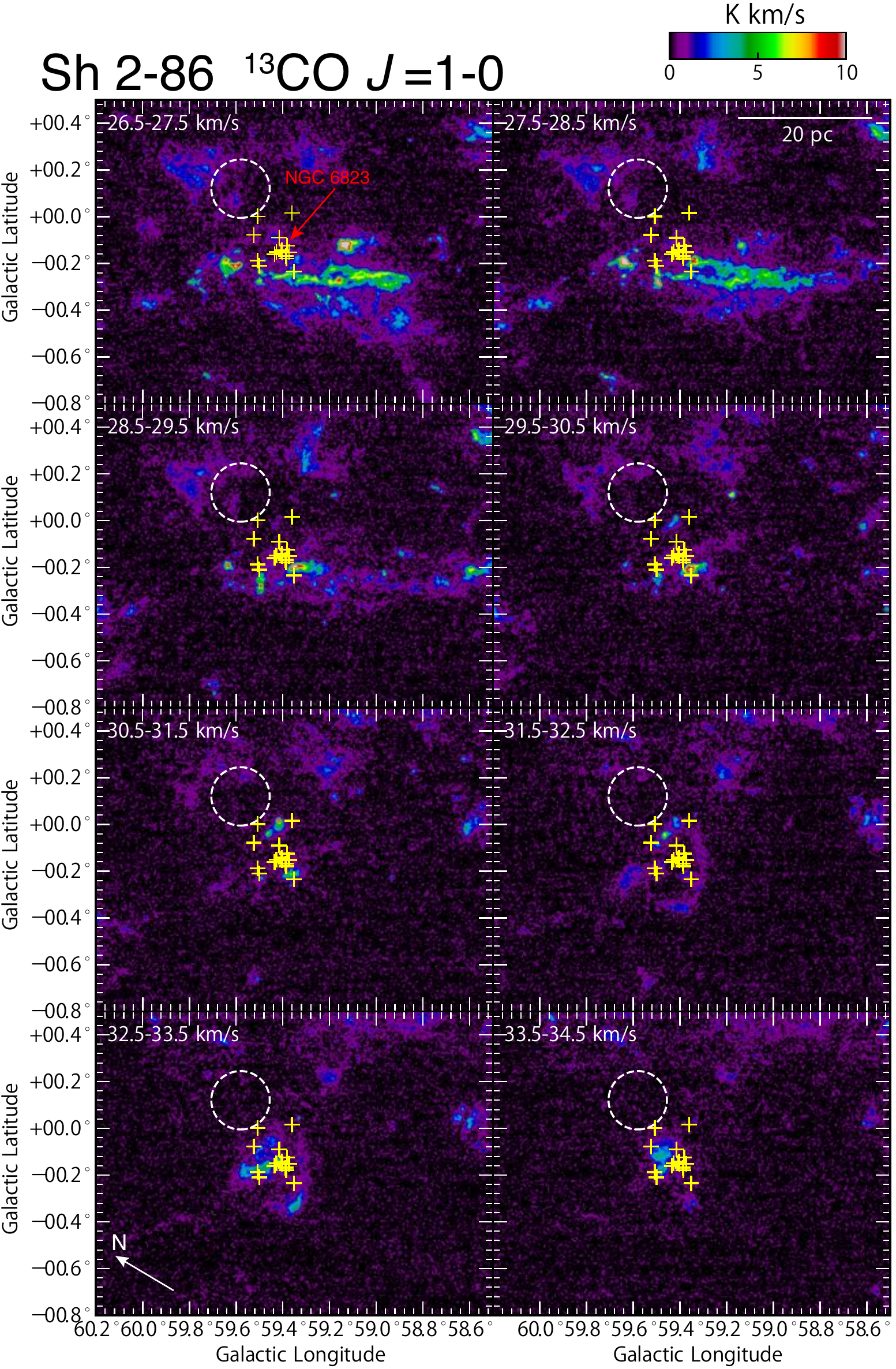}
\end{center}
\caption{{(Continued.)}}
\label{13COch2}
\end{figure*}

\begin{figure*}[h]
\begin{center} 
 \includegraphics[width=14cm]{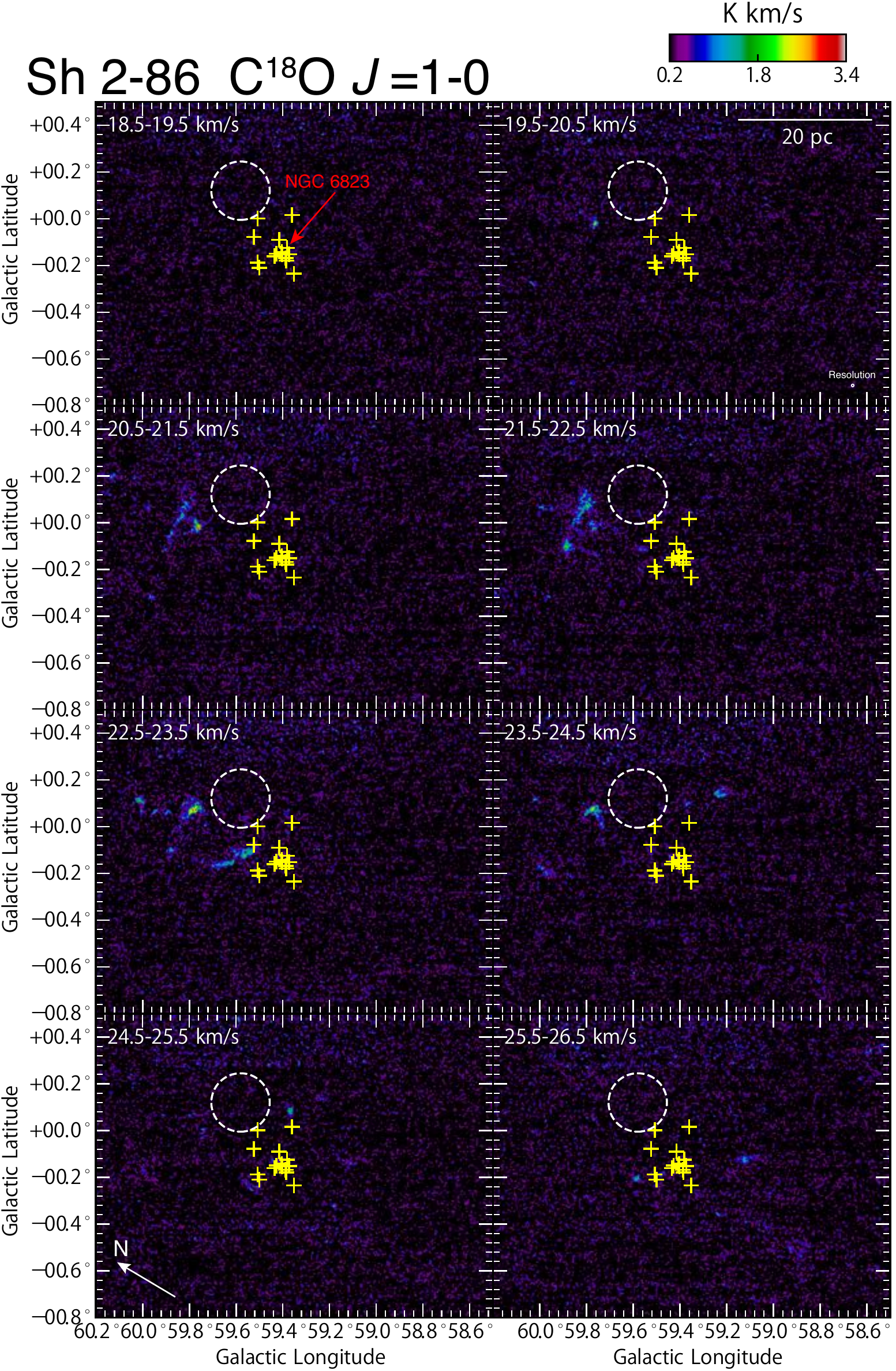}
\end{center}
\caption{Same as Figure\ref{13COch1}, but for C$^{18}$O $J=$ 1--0}
\label{C18Och1}
\end{figure*}
\addtocounter{figure}{-1}

\begin{figure*}[h]
\begin{center} 
 \includegraphics[width=14cm]{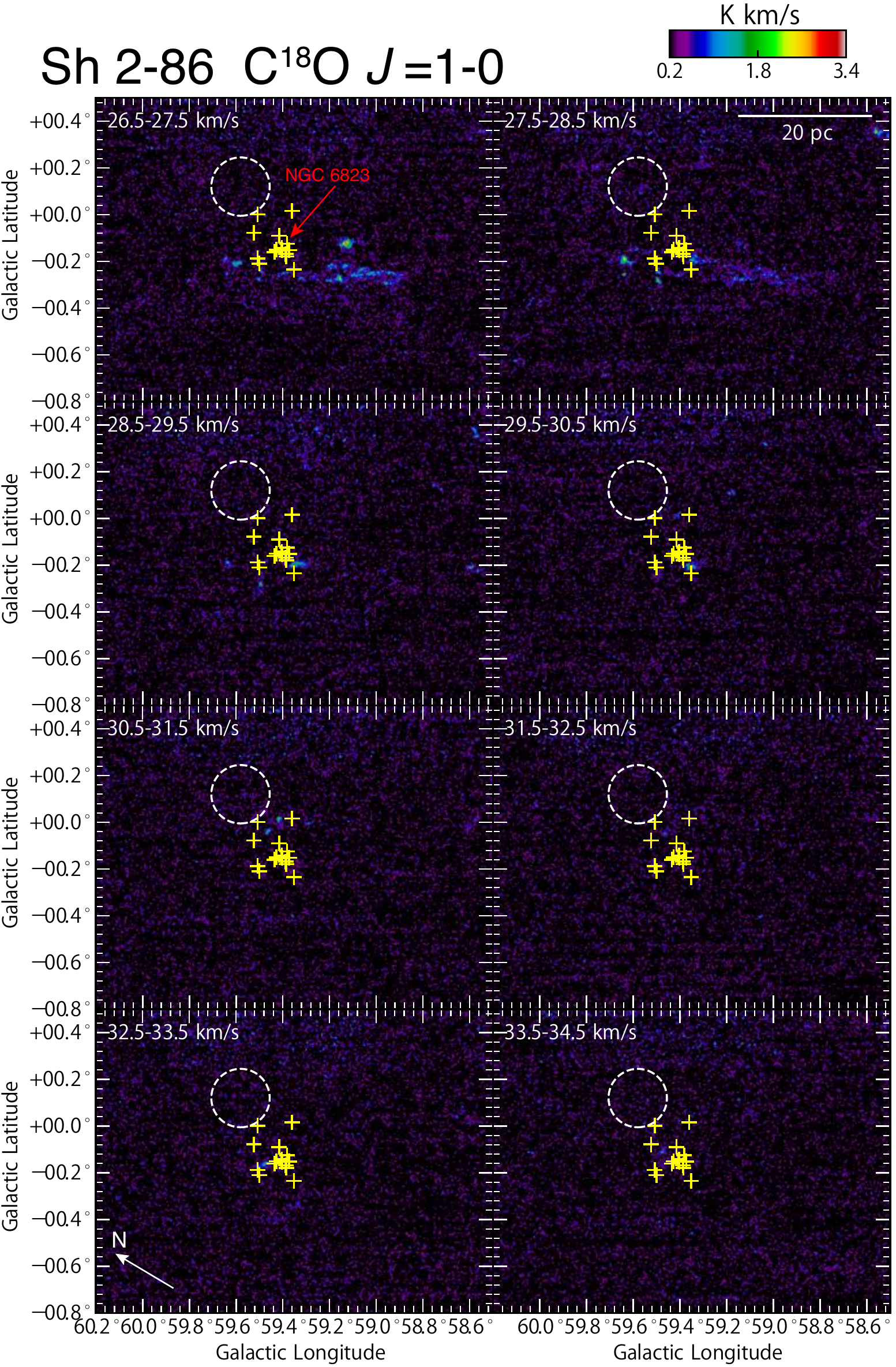}
\end{center}
\caption{{(Continued.)}}
\label{C18Och2}
\end{figure*}



\begin{thebibliography}{}
\bibitem[Abreu-Vicente et al.(2016)]{2016A&A...590A.131A} Abreu-Vicente, J., Ragan, S., Kainulainen, J., et al.\ 2016, \aap, 590, A131. doi:10.1051/0004-6361/201527674
\bibitem[Astropy Collaboration et al.(2013)]{2013A&A...558A..33A} Astropy Collaboration, Robitaille, T.~P., Tollerud, E.~J., et al.\ 2013, \aap, 558, A33
\bibitem[Astropy Collaboration et al.(2018)]{2018AJ....156..123A} Astropy Collaboration, Price-Whelan, A.~M., Sip{\H{o}}cz, B.~M., et al.\ 2018, \aj, 156, 123
\bibitem[Arzoumanian et al.(2018)]{2018PASJ...70...96A} {Arzoumanian, D., Shimajiri, Y., Inutsuka, S.-. ichiro ., et al.\ 2018, \pasj, 70, 96. doi:10.1093/pasj/psy095}
\bibitem[Barkhatova(1957)]{1957SvA.....1..822B} Barkhatova, K.~A.\ 1957, \sovast, 1, 822
\bibitem[Barsony(1989)]{1989ApJ...345..268B} Barsony, M.\ 1989, \apj, 345, 268
\bibitem[Berriman \& Good(2017)]{2017PASP..129e8006B} Berriman, G.~B. \& Good, J.~C.\ 2017, \pasp, 129, 058006. doi:10.1088/1538-3873/aa5456
\bibitem[Beuther et al.(2004)]{2004ApJ...615..832B} Beuther, H., Schilke, P., \& Wyrowski, F.\ 2004, \apj, 615, 832
\bibitem[Bica et al.(2008)]{2008A&A...489.1129B} Bica, E., Bonatto, C., \& Dutra, C.~M.\ 2008, \aap, 489, 1129
\bibitem[Billot et al.(2010)]{2010ApJ...712..797B} Billot, N., Noriega-Crespo, A., Carey, S., et al.\ 2010, \apj, 712, 797
\bibitem[Blitz et al.(2007)]{2007prpl.conf...81B} Blitz, L., Fukui, Y., Kawamura, A., et al.\ 2007, Protostars and Planets V, 81
\bibitem[Bolatto et al.(2013)]{2013ARA&A..51..207B} Bolatto, A.~D., Wolfire, M., \& Leroy, A.~K.\ 2013, \araa, 51, 20
\bibitem[Burton(1970)]{1970A&AS....2..291B} Burton, W.~B.\ 1970, \aaps, 2, 291
\bibitem[Burton \& Gordon(1978)]{1978A&A....63....7B} Burton, W.~B. \& Gordon, M.~A.\ 1978, \aap, 63, 7
\bibitem[Cappa et al.(2002)]{2002A&A...395..955C} Cappa, C., Pineault, S., Arnal, E.~M., et al.\ 2002, \aap, 395, 955
\bibitem[Carpenter \& Sanders(1998)]{1998AJ....116.1856C} Carpenter, J.~M. \& Sanders, D.~B.\ 1998, \aj, 116, 1856
\bibitem[Chapin et al.(2008)]{2008ApJ...681..428C} Chapin, E.~L., Ade, P.~A.~R., Bock, J.~J., et al.\ 2008, \apj, 681, 428
\bibitem[Chavarr{\'\i}a et al.(2008)]{2008ApJ...682..445C} {Chavarr{\'\i}a, L.~A., Allen, L.~E., Hora, J.~L., et al.\ 2008, \apj, 682, 445. doi:10.1086/588810}
\bibitem[Chen et al.(2003)]{2003A&A...401..185C} Chen, Y., Zheng, X.-W., Yao, Y., et al.\ 2003, \aap, 401, 185
\bibitem[Cohen et al.(1980)]{1980ApJ...239L..53C} Cohen, R.~S., Cong, H., Dame, T.~M., et al.\ 1980, \apjl, 239, L53
\bibitem[Collinder(1931)]{1931AnLun...2....1C} Collinder, P.\ 1931, Annals of the Observatory of Lund, 2, B1
\bibitem[Combes(1991)]{1991ARA&A..29..195C} Combes, F.\ 1991, \araa, 29, 195
\bibitem[Cosentino et al.(2020)]{2020MNRAS.499.1666C} {Cosentino, G., Jim{\'e}nez-Serra, I., Henshaw, J.~D., et al.\ 2020, \mnras, 499, 1666. doi:10.1093/mnras/staa2942}
\bibitem[Dame, \& Thaddeus(1985)]{1985ApJ...297..751D} Dame, T.~M., \& Thaddeus, P.\ 1985, \apj, 297, 751
\bibitem[Dame et al.(1986)]{1986ApJ...305..892D} Dame, T.~M., Elmegreen, B.~G., Cohen, R.~S., et al.\ 1986, \apj, 305, 892
\bibitem[Dame et al.(1987)]{1987ApJ...322..706D} Dame, T.~M., Ungerechts, H., Cohen, R.~S., et al.\ 1987, \apj, 322, 706
\bibitem[Deharveng, \& Maucherat(1978)]{1978A&A....70...19D} Deharveng, L., \& Maucherat, M.\ 1978, \aap, 70, 19
\bibitem[Deharveng et al.(2000)]{2000A&A...360.1107D} Deharveng, L., Nadeau, D., Zavagno, A., et al.\ 2000, \aap, 360, 1107
\bibitem[Deharveng et al.(2010)]{2010A&A...523A...6D} {Deharveng, L., Schuller, F., Anderson, L.~D., et al.\ 2010, \aap, 523, A6. doi:10.1051/0004-6361/201014422}
\bibitem[Dobashi et al.(2014)]{2014ApJ...797...58D} {Dobashi, K., Matsumoto, T., Shimoikura, T., et al.\ 2014, \apj, 797, 58. doi:10.1088/0004-637X/797/1/58}
\bibitem[Dobbs \& Bonnell(2006)]{2006MNRAS.367..873D} Dobbs, C.~L. \& Bonnell, I.~A.\ 2006, \mnras, 367, 873
\bibitem[Dobbs et al.(2014)]{2014prpl.conf....3D} Dobbs, C.~L., Krumholz, M.~R., Ballesteros-Paredes, J., et al.\ 2014, Protostars and Planets VI, 3
\bibitem[Druard et al.(2014)]{2014A&A...567A.118D} {Druard, C., Braine, J., Schuster, K.~F., et al.\ 2014, \aap, 567, A118. doi:10.1051/0004-6361/201423682}
\bibitem[Duarte-Cabral \& Dobbs(2017)]{2017MNRAS.470.4261D} Duarte-Cabral, A. \& Dobbs, C.~L.\ 2017, \mnras, 470, 4261. doi:10.1093/mnras/stx1524
 \bibitem[Evans et al.(1981)]{1981ApJ...250..200E} Evans, N.~J., Blair, G.~N., Harvey, P., et al.\ 1981, \apj, 250, 200
\bibitem[Ehlerov{\'a} et al.(2001)]{2001A&A...374..682E} Ehlerov{\'a}, S., Palou{\v{s}}, J., \& Huchtmeier, W.~K.\ 2001, \aap, 374, 682
\bibitem[Elmegreen, \& Lada(1977)]{1977ApJ...214..725E} Elmegreen, B.~G., \& Lada, C.~J.\ 1977, \apj, 214, 725
\bibitem[Erickson(1971)]{1971A&A....10..270E} Erickson, R.~R.\ 1971, \aap, 10, 270
\bibitem[Frerking et al.(1982)]{1982ApJ...262..590F} Frerking, M.~A., Langer, W.~D., \& Wilson, R.~W.\ 1982, \apj, 262, 590
\bibitem[Fujita et al.(2019)]{2019ApJ...872...49F} Fujita, S., Torii, K., Tachihara, K., et al.\ 2019, \apj, 872, 49
\bibitem[Fujita et al.(2021a)]{2019PASJ..tmp...46F} Fujita, S., Torii, K., Kuno, N., et al.\ 2021a, \pasj, 73, S172. doi:10.1093/pasj/psz028
\bibitem[Fujita et al.(2021b)]{2020PASJ..tmp..238F} Fujita, S., Sano, H., Enokiya, R., et al.\ 2021b, \pasj, 73, S201. doi:10.1093/pasj/psaa078

\bibitem[Fukui, \& Kawamura(2010)]{2010ARA&A..48..547F} Fukui, Y., \& Kawamura, A.\ 2010, \araa, 48, 547
\bibitem[Fukui et al.(2015)]{2015ApJ...807L...4F} Fukui, Y., Harada, R., Tokuda, K., et al.\ 2015, \apjl, 807, L4
\bibitem[Fukui et al.(2016)]{2016ApJ...820...26F} {Fukui, Y., Torii, K., Ohama, A., et al.\ 2016, \apj, 820, 26. doi:10.3847/0004-637X/820/1/26}
\bibitem[Fukui et al.(2019)]{2019ApJ...886...14F} Fukui, Y., Tokuda, K., Saigo, K., et al.\ 2019, \apj, 886, 14
\bibitem[Fukui et al.(2021)]{Fukui+2020} Fukui, Y., Habe, A., Inoue, T., et al.\ 2021, \pasj, 73, S1. doi:10.1093/pasj/psaa103
\bibitem[Goldreich \& Kwan(1974)]{1974ApJ...189..441G} Goldreich, P. \& Kwan, J.\ 1974, \apj, 189, 441. doi:10.1086/152821
\bibitem[Goodman et al.(2014)]{2014ApJ...797...53G} Goodman, A.~A., Alves, J., Beaumont, C.~N., et al.\ 2014, \apj, 797, 53. doi:10.1088/0004-637X/797/1/53
\bibitem[Gratier et al.(2010)]{2010A&A...522A...3G} {Gratier, P., Braine, J., Rodriguez-Fernandez, N.~J., et al.\ 2010, \aap, 522, A3. doi:10.1051/0004-6361/201014441}
\bibitem[Griffin et al.(2010)]{2010A&A...518L...3G} Griffin, M.~J., Abergel, A., Abreu, A., et al.\ 2010, \aap, 518, L3
\bibitem[Green(2019)]{2019JApA...40...36G} Green, D.~A.\ 2019, Journal of Astrophysics and Astronomy, 40, 36
\bibitem[Guetter(1992)]{1992AJ....103..197G} Guetter, H.~H.\ 1992, \aj, 103, 197
\bibitem[Heyer, \& Dame(2015)]{2015ARA&A..53..583H} Heyer, M., \& Dame, T.~M.\ 2015, \araa, 53, 583
\bibitem[Habe \& Ohta(1992)]{1992PASJ...44..203H} Habe, A. \& Ohta, K.\ 1992, \pasj, 44, 203
\bibitem[Hasegawa et al.(2017)]{2017PASJ...69...91H} Hasegawa, Y., Asayama, S., Harada, R., et al.\ 2017, \pasj, 69, 91. doi:10.1093/pasj/psx098
\bibitem[Haworth et al.(2015)]{2015MNRAS.450...10H} {Haworth, T.~J., Tasker, E.~J., Fukui, Y., et al.\ 2015, \mnras, 450, 10. doi:10.1093/mnras/stv639}
\bibitem[Higuchi et al.(2009)]{2009ApJ...705..468H} Higuchi, A.~E., Kurono, Y., Saito, M., et al.\ 2009, \apj, 705, 468
\bibitem[Higuchi et al.(2010)]{2010ApJ...719.1813H} Higuchi, A.~E., Kurono, Y., Saito, M., et al.\ 2010, \apj, 719, 1813
\bibitem[Hunter(2007)]{2007CSE.....9...90H} Hunter, J.~D.\ 2007, Computing in Science and Engineering, 9, 90
\bibitem[Hasegawa et al.(1994)]{1994ApJ...429L..77H} {Hasegawa, T., Sato, F., Whiteoak, J.~B., et al.\ 1994, \apjl, 429, L77. doi:10.1086/187417}
\bibitem[Inoue \& Fukui(2013)]{2013ApJ...774L..31I} Inoue, T. \& Fukui, Y.\ 2013, \apjl, 774, L31. doi:10.1088/2041-8205/774/2/L31
\bibitem[Inoue et al.(2018)]{2018PASJ...70S..53I} Inoue, T., Hennebelle, P., Fukui, Y., et al.\ 2018, \pasj, 70, S53. doi:10.1093/pasj/psx089
\bibitem[Jackson et al.(2010)]{2010ApJ...719L.185J} Jackson, J.~M., Finn, S.~C., Chambers, E.~T., et al.\ 2010, \apjl, 719, L185. doi:10.1088/2041-8205/719/2/L185
\bibitem[Kamazaki et al.(2012)]{2012PASJ...64...29K} Kamazaki, T., Okumura, S.~K., Chikada, Y., et al.\ 2012, \pasj, 64, 29
\bibitem[Kauffmann et al.(2008)]{2008A&A...487..993K} Kauffmann, J., Bertoldi, F., Bourke, T.~L., et al.\ 2008, \aap, 487, 993
\bibitem[Koda et al.(2009)]{2009ApJ...700L.132K} Koda, J., Scoville, N., Sawada, T., et al.\ 2009, \apjl, 700, L132
\bibitem[Kohno et al.(2018)]{2018PASJ...70S..50K} Kohno, M., Torii, K., Tachihara, K., et al.\ 2018, \pasj, 70, S50
\bibitem[Kohno et al.(2021)]{2020arXiv200110693K} Kohno, M., Tachihara, K., Torii, K., et al.\ 2021a, \pasj, 73, S129. doi:10.1093/pasj/psaa015
\bibitem[Kohno et al.(2021)]{2021PASJ...73S.338K} {Kohno, M., Tachihara, K., Fujita, S., et al.\ 2021b, \pasj, 73, S338. doi:10.1093/pasj/psy109}
\bibitem[Kondo et al.(2021)]{2021arXiv210301610K} {Kondo, H., Tokuda, K., Muraoka, K., et al.\ 2021, \apj, 912, 66. doi:10.3847/1538-4357/abeb65}
\bibitem[Koo(1999)]{1999ApJ...518..760K} Koo, B.-C.\ 1999, \apj, 518, 760
\bibitem[Kumar et al.(2004)]{2004MNRAS.353..991K} Kumar, B., Sagar, R., Sanwal, B.~B., et al.\ 2004, \mnras, 353, 991
\bibitem[Kumar et al.(2004)]{2004MNRAS.353.1025K} Kumar, M.~S.~N., Kamath, U.~S., \& Davis, C.~J.\ 2004, \mnras, 353, 1025
\bibitem[Kuno et al.(2011)]{Proc..2011} Kuno, N., et al. 2011, in Proc. 2011 XXXth URSI General Assembly
and Scientific Symposium (New York: IEEE), 3670 \footnote{\url{http://ieeexplore.ieee.org/xpl/articleDetails.jsp?arnumber=6051296}}
\bibitem[Kutner, \& Ulich(1981)]{1981ApJ...250..341K} Kutner, M.~L., \& Ulich, B.~L.\ 1981, \apj, 250, 341
\bibitem[Lada, \& Lada(2003)]{2003ARA&A..41...57L} Lada, C.~J., \& Lada, E.~A.\ 2003, \araa, 41, 57
\bibitem[Lada et al.(2003)]{2003ApJ...586..286L} {Lada, C.~J., Bergin, E.~A., Alves, J.~F., et al.\ 2003, \apj, 586, 286. doi:10.1086/367610}
\bibitem[Li et al.(2013)]{2013A&A...559A..34L} Li, G.-X., Wyrowski, F., Menten, K., et al.\ 2013, \aap, 559, A34. doi:10.1051/0004-6361/201322411
\bibitem[Liu et al.(2019)]{2019MNRAS.487.1259L} {Liu, H.-L., Stutz, A., \& Yuan, J.-H.\ 2019, \mnras, 487, 1259. doi:10.1093/mnras/stz1340}
\bibitem[Lortet-Zuckermann(1974)]{1974A&A....30...67L} Lortet-Zuckermann, M.~C.\ 1974, \aap, 30, 67
\bibitem[Mar{\'\i}n-Franch et al.(2009)]{2009A&A...502..559M} Mar{\'\i}n-Franch, A., Herrero, A., Lenorzer, A., et al.\ 2009, \aap, 502, 559
\bibitem[Mart{\'\i}n-Hern{\'a}ndez et al.(2008)]{2008A&A...489..229M} Mart{\'\i}n-Hern{\'a}ndez, N.~L., Bik, A., Puga, E., et al.\ 2008, \aap, 489, 229
\bibitem[Massey et al.(1995)]{1995ApJ...454..151M} Massey, P., Johnson, K.~E., \& Degioia-Eastwood, K.\ 1995, \apj, 454, 151
\bibitem[Matsumoto et al.(2015)]{2015ApJ...801...77M} {Matsumoto, T., Dobashi, K., \& Shimoikura, T.\ 2015, \apj, 801, 77. doi:10.1088/0004-637X/801/2/77}
\bibitem[McClure-Griffiths, \& Dickey(2016)]{2016ApJ...831..124M} McClure-Griffiths, N.~M., \& Dickey, J.~M.\ 2016, \apj, 831, 124
\bibitem[McKee, \& Ostriker(2007)]{2007ARA&A..45..565M} McKee, C.~F., \& Ostriker, E.~C.\ 2007, \araa, 45, 565
\bibitem[Minamidani et al.(2015)]{2015EAS....75..193M} Minamidani, T., Umemoto, T., Nishimura, A., et al.\ 2015, EAS Publications Series, 75-76, 193. doi:10.1051/eas/1575036
\bibitem[Minamidani et al.(2016)]{2016SPIE.9914E..1ZM} Minamidani, T., Nishimura, A., Miyamoto, Y., et al.\ 2016, \procspie, 9914, 99141Z. doi:10.1117/12.2232137
\bibitem[Miyamoto et al.(2014)]{2014PASJ...66...36M} Miyamoto, Y., Nakai, N., \& Kuno, N.\ 2014, \pasj, 66, 36
\bibitem[Molinari et al.(2010a)]{2010PASP..122..314M} Molinari, S., Swinyard, B., Bally, J., et al.\ 2010a, \pasp, 122, 314
\bibitem[Molinari et al.(2010b)]{2010A&A...518L.100M} Molinari, S., Swinyard, B., Bally, J., et al.\ 2010b, \aap, 518, L100
\bibitem[Molinari et al.(2016)]{2016A&A...591A.149M} Molinari, S., Schisano, E., Elia, D., et al.\ 2016, \aap, 591, A149
\bibitem[Morgan et al.(1953)]{1953ApJ...118..318M} Morgan, W.~W., Whitford, A.~E., \& Code, A.~D.\ 1953, \apj, 118, 318
\bibitem[Motte et al.(2018)]{2018ARA&A..56...41M} Motte, F., Bontemps, S., \& Louvet, F.\ 2018, \araa, 56, 41
\bibitem[Mottram \& Brunt(2012)]{2012MNRAS.420...10M} Mottram, J.~C. \& Brunt, C.~M.\ 2012, \mnras, 420, 10. doi:10.1111/j.1365-2966.2011.19843.x
\bibitem[Nakajima et al.(2019)]{2019PASJ...71S..17N} Nakajima, T., Inoue, H., Fujii, Y., et al.\ 2019, \pasj, 71, S17
\bibitem[Nakamura et al.(2014)]{2014ApJ...791L..23N} {Nakamura, F., Sugitani, K., Tanaka, T., et al.\ 2014, \apjl, 791, L23. doi:10.1088/2041-8205/791/2/L23}
\bibitem[Nakanishi et al.(2020)]{2020PASJ...72...43N} {Nakanishi, H., Fujita, S., Tachihara, K., et al.\ 2020, \pasj, 72, 43. doi:10.1093/pasj/psaa027}
\bibitem[Nishimura et al.(2015)]{2015ApJS..216...18N} Nishimura, A., Tokuda, K., Kimura, K., et al.\ 2015, \apjs, 216, 18
\bibitem[Nishimura et al.(2018)]{2018PASJ...70S..42N} Nishimura, A., Minamidani, T., Umemoto, T., et al.\ 2018, \pasj, 70, S42
\bibitem[Nishimura et al.(2021)]{2020arXiv200805939N} Nishimura, A., Fujita, S., Kohno, M., et al.\ 2021, \pasj, 73, S285. doi:10.1093/pasj/psaa083

\bibitem[Nishimura et al.(2020)]{2020arXiv201200906N} Nishimura, A., Tokuda, K., Harada, R., et al.\ 2020, \procspie, 11445, 114457F. doi:10.1117/12.2560955
\bibitem[Ohashi et al.(1997)]{1997ApJ...475..211O} {Ohashi, N., Hayashi, M., Ho, P.~T.~P., et al.\ 1997, \apj, 475, 211. doi:10.1086/303533}
\bibitem[Olmi et al.(2010)]{2010ApJ...715.1132O} Olmi, L., Araya, E.~D., Chapin, E.~L., et al.\ 2010, \apj, 715, 1132. doi:10.1088/0004-637X/715/2/1132
\bibitem[Onishi et al.(2013)]{2013PASJ...65...78O} Onishi, T., Nishimura, A., Ota, Y., et al.\ 2013, \pasj, 65, 78
\bibitem[Perez, \& Granger(2007)]{2007CSE.....9c..21P} Perez, F., \& Granger, B.~E.\ 2007, Computing in Science and Engineering, 9, 21
\bibitem[Pigulski et al.(2000)]{2000AcA....50..113P} Pigulski, A., Kolaczkowski, Z., \& Kopacki, G.\ 2000, ACTAA, 50, 113
\bibitem[Pilbratt et al.(2010)]{2010A&A...518L...1P} Pilbratt, G.~L., Riedinger, J.~R., Passvogel, T., et al.\ 2010, \aap, 518, L1
\bibitem[Pineda et al.(2008)]{2008ApJ...679..481P} Pineda, J.~E., Caselli, P., \& Goodman, A.~A.\ 2008, \apj, 679, 481
\bibitem[Pineda et al.(2010)]{2010ApJ...721..686P} Pineda, J.~L., Goldsmith, P.~F., Chapman, N., et al.\ 2010, \apj, 721, 686
\bibitem[Pipher et al.(1977)]{1977A&A....59..215P} Pipher, J.~L., Sharpless, S., Savedoff, M.~P., et al.\ 1977, \aap, 59, 215
\bibitem[Poglitsch et al.(2010)]{2010A&A...518L...2P} Poglitsch, A., Waelkens, C., Geis, N., et al.\ 2010, \aap, 518, L2
\bibitem[Priestley \& Whitworth(2021)]{2021MNRAS.506..775P} {Priestley, F.~D. \& Whitworth, A.~P.\ 2021, \mnras, 506, 775. doi:10.1093/mnras/stab1777}
\bibitem[Ragan et al.(2014)]{2014A&A...568A..73R} Ragan, S.~E., Henning, T., Tackenberg, J., et al.\ 2014, \aap, 568, A73
\bibitem[Reid et al.(2016)]{2016ApJ...823...77R} Reid, M.~J., Dame, T.~M., Menten, K.~M., et al.\ 2016, \apj, 823, 77
\bibitem[Reid et al.(2019)]{2019ApJ...885..131R} Reid, M.~J., Menten, K.~M., Brunthaler, A., et al.\ 2019, \apj, 885, 131
\bibitem[Riaz et al.(2012)]{2012MNRAS.419.1887R} Riaz, B., Mart{\'\i}n, E.~L., Tata, R., et al.\ 2012, \mnras, 419, 1887
\bibitem[Robitaille \& Bressert(2012)]{2012ascl.soft08017R} Robitaille, T., \& Bressert, E.\ 2012, APLpy: Astronomical Plotting Library in Python, ascl:1208.017
\bibitem[Rod{\'o}n et al.(2012)]{2012A&A...545A..51R} Rod{\'o}n, J.~A., Beuther, H., \& Schilke, P.\ 2012, \aap, 545, A51
\bibitem[Russeil et al.(2013)]{2013A&A...554A..42R} Russeil, D., Schneider, N., Anderson, L.~D., et al.\ 2013, \aap, 554, A42
\bibitem[Sagar, \& Joshi(1981)]{1981Ap&SS..75..465S} Sagar, R., \& Joshi, U.~C.\ 1981, \apss, 75, 465
\bibitem[Saigo et al.(2017)]{2017ApJ...835..108S} Saigo, K., Onishi, T., Nayak, O., et al.\ 2017, \apj, 835, 108
\bibitem[Saito et al.(2007)]{2007ApJ...659..459S} Saito, H., Saito, M., Sunada, K., et al.\ 2007, \apj, 659, 459
\bibitem[Sano et al.(2021)]{2021PASJ...73S..62S} Sano, H., Tsuge, K., Tokuda, K., et al.\ 2021, \pasj, 73, S62. doi:10.1093/pasj/psaa045
\bibitem[Sault et al.(1995)]{1995ASPC...77..433S} {Sault, R.~J., Teuben, P.~J., \& Wright, M.~C.~H.\ 1995, Astronomical Data Analysis Software and Systems IV, 77, 433}
\bibitem[Sawada et al.(2008)]{2008PASJ...60..445S} Sawada, T., Ikeda, N., Sunada, K., et al.\ 2008, \pasj, 60, 445
\bibitem[Sharpless(1959)]{1959ApJS....4..257S} Sharpless, S.\ 1959, \apjs, 4, 257
\bibitem[Shi, \& Hu(1999)]{1999A&AS..136..313S} Shi, H.~M., \& Hu, J.~Y.\ 1999, \aaps, 136, 313
\bibitem[Shimoikura et al.(2016)]{2016ApJ...832..205S} {Shimoikura, T., Dobashi, K., Matsumoto, T., et al.\ 2016, \apj, 832, 205. doi:10.3847/0004-637X/832/2/205}
\bibitem[Shimoikura et al.(2018)]{2018ApJ...855...45S} {Shimoikura, T., Dobashi, K., Nakamura, F., et al.\ 2018, \apj, 855, 45. doi:10.3847/1538-4357/aaaccd}
\bibitem[Sofue et al.(2019)]{2019PASJ...71S...1S} {Sofue, Y., Kohno, M., Torii, K., et al.\ 2019, \pasj, 71, S1. doi:10.1093/pasj/psy094}
\bibitem[Sofue \& Kohno(2020)]{2020MNRAS} Sofue, Y. \& Kohno, M.\ 2020, \mnras, 497, 1851. doi:10.1093/mnras/staa2056
\bibitem[Stone(1988)]{1988AJ.....96.1389S} Stone, R.~C.\ 1988, \aj, 96, 1389
\bibitem[Szymczak et al.(2000)]{2000A&AS..143..269S} Szymczak, M., Hrynek, G., \& Kus, A.~J.\ 2000, \aaps, 143, 269. doi:10.1051/aas:2000334
\bibitem[Takahira et al.(2014)]{2014ApJ...792...63T} {Takahira, K., Tasker, E.~J., \& Habe, A.\ 2014, \apj, 792, 63. doi:10.1088/0004-637X/792/1/63}

\bibitem[Taylor et al.(1992)]{1992AJ....103..931T} Taylor, A.~R., Wallace, B.~J., \& Goss, W.~M.\ 1992, \aj, 103, 931
\bibitem[Tokuda et al.(2019)]{2019ApJ...886...15T} Tokuda, K., Fukui, Y., Harada, R., et al.\ 2019, \apj, 886, 15
\bibitem[Tokuda et al.(2020)]{2020ApJ...896...36T} Tokuda, K., Muraoka, K., Kondo, H., et al.\ 2020, \apj, 896, 36
\bibitem[Torii et al.(2011)]{2011ApJ...738...46T} {Torii, K., Enokiya, R., Sano, H., et al.\ 2011, \apj, 738, 46. doi:10.1088/0004-637X/738/1/46}
\bibitem[Torii et al.(2018)]{2018PASJ...70S..51T} Torii, K., Fujita, S., Matsuo, M., et al.\ 2018, \pasj, 70, S51. doi:10.1093/pasj/psy019
\bibitem[Torii et al.(2019)]{2019PASJ...71S...2T} Torii, K., Fujita, S., Nishimura, A., et al.\ 2019, \pasj, 71, S2
\bibitem[Torii et al.(2021)]{2018PASJ..tmp..121T} Torii, K., Hattori, Y., Matsuo, M., et al.\ 2021, \pasj, 73, S368. doi:10.1093/pasj/psy098
\bibitem[Tsuge et al.(2021)]{2021PASJ...73S..35T} {Tsuge, K., Fukui, Y., Tachihara, K., et al.\ 2021, \pasj, 73, S35. doi:10.1093/pasj/psaa033}
\bibitem[Tsuge et al.(2021)]{2021PASJ...73..417T} {Tsuge, K., Tachihara, K., Fukui, Y., et al.\ 2021, \pasj, 73, 417. doi:10.1093/pasj/psab008}
\bibitem[Turner(1979)]{1979JRASC..73...74T} Turner, D.~G.\ 1979, \jrasc, 73, 74
\bibitem[Turner(1986)]{1986A&A...167..157T} Turner, D.~G.\ 1986, \aap, 167, 157
\bibitem[Ulich, \& Haas(1976)]{1976ApJS...30..247U} Ulich, B.~L., \& Haas, R.~W.\ 1976, \apjs, 30, 247
\bibitem[Umemoto et al.(2017)]{2017PASJ...69...78U} Umemoto, T., Minamidani, T., Kuno, N., et al.\ 2017, \pasj, 69, 78
\bibitem[van der Walt et al.(2011)]{2011CSE....13b..22V} van der Walt, S., Colbert, S.~C., \& Varoquaux, G.\ 2011, Computing in Science and Engineering, 13, 22
\bibitem[VERA Collaboration et al.(2020)]{2020arXiv200203089V} VERA Collaboration, Hirota, T., Nagayama, T., et al.\ 2020, \pasj, 72, 50. doi:10.1093/pasj/psaa018
\bibitem[Wada \& Koda(2004)]{2004MNRAS.349..270W} Wada, K., \& Koda, J.\ 2004, \mnras, 349, 270
\bibitem[Wang et al.(2015)]{2015MNRAS.450.4043W} Wang, K., Testi, L., Ginsburg, A., et al.\ 2015, \mnras, 450, 4043. doi:10.1093/mnras/stv735
\bibitem[Wang et al.(2020)]{2020A&A...641A..53W} Wang, Y., Beuther, H., Schneider, N., et al.\ 2020, \aap, 641, A53. doi:10.1051/0004-6361/202037928
\bibitem[Wilson, \& Rood(1994)]{1994ARA&A..32..191W} Wilson, T.~L., \& Rood, R.\ 1994, \araa, 32, 191
\bibitem[Wilson et al.(2013)]{2013tra..book.....W} Wilson, T.~L., Rohlfs, K., \& H{\"u}ttemeister, S.\ 2013, Tools of Radio Astronomy; Astronomy and Astrophysics Library. ISBN 978-3-642-39949-7. Springer-Verlag Berlin Heidelberg
\bibitem[Wu et al.(2015)]{2015ApJ...811...56W} {Wu, B., Van Loo, S., Tan, J.~C., et al.\ 2015, \apj, 811, 56. doi:10.1088/0004-637X/811/1/56}
\bibitem[Xu et al.(2005)]{2005ChJAA...5..165X} Xu, J.-W., Zhang, X.-Z., \& Han, J.-L.\ 2005, CIAA, 5, 165
\bibitem[Xu, \& Wang(2012)]{2012A&A...543A..24X} Xu, J.-L., \& Wang, J.-J.\ 2012, \aap, 543, A24
\bibitem[Xu et al.(2009)]{2009ApJ...693..413X} Xu, Y., Reid, M.~J., Menten, K.~M., et al.\ 2009, \apj, 693, 413
\bibitem[Xu \& Ju(2014)]{2014A&A...569A..36X} {Xu, J.-L. \& Ju, B.-G.\ 2014, \aap, 569, A36. doi:10.1051/0004-6361/201423952}
\bibitem[Xu et al.(2016)]{2016SciA....2E0878X} Xu, Y., Reid, M., Dame, T., et al.\ 2016, Science Advances, 2, e1600878
\bibitem[Xu et al.(2018)]{2018A&A...616L..15X} Xu, Y., Bian, S.~B., Reid, M.~J., et al.\ 2018, \aap, 616, L15
\bibitem[Xu et al.(2021)]{2021arXiv210100158X} Xu, Y., Hou, L.~G., Bian, S.~B., et al.\ 2021, \aap, 645, L8. doi:10.1051/0004-6361/202040103
\bibitem[Xue, \& Wu(2008)]{2008ApJ...680..446X} Xue, R., \& Wu, Y.\ 2008, \apj, 680, 446
\bibitem[Yajima et al.(2021)]{2020arXiv201208523Y} Yajima, Y., Sorai, K., Miyamoto, Y., et al.\ 2021, \pasj. doi:10.1093/pasj/psaa119
\bibitem[Yamada et al.(2021)]{2021arXiv210601852Y} {Yamada, R., Fukui, Y., Sano, H., et al.\ 2021, arXiv:2106.01852}
\bibitem[Yoda et al.(2010)]{2010PASJ...62.1277Y} Yoda, T., Handa, T., Kohno, K., et al.\ 2010, \pasj, 62, 1277
\bibitem[Zhang et al.(2019)]{2019A&A...622A..52Z} Zhang, M., Kainulainen, J., Mattern, M., et al.\ 2019, \aap, 622, A52. doi:10.1051/0004-6361/201732400
\bibitem[Zucker et al.(2015)]{2015ApJ...815...23Z} Zucker, C., Battersby, C., \& Goodman, A.\ 2015, \apj, 815, 23. doi:10.1088/0004-637X/815/1/23
\bibitem[Zychov{\'a} \& Ehlerov{\'a}(2016)]{2016A&A...595A..49Z} {Zychov{\'a}, L. \& Ehlerov{\'a}, S.\ 2016, \aap, 595, A49. doi:10.1051/0004-6361/201527897}
\end{thebibliography}
\end{document}